\documentclass[11pt, a4paper]{article}
\usepackage[utf8]{inputenc}
\usepackage[T1]{fontenc}
\usepackage[english]{babel}
\usepackage{csquotes}
\usepackage[backend=biber, natbib=true, style=apa]{biblatex}
\usepackage{amsmath, amsfonts, amssymb}
\usepackage{graphicx} 
\usepackage{caption}
\usepackage[belowskip=2ex]{subcaption}
\usepackage[margin=1in]{geometry}
\usepackage{verbatim}
\usepackage{float}
\usepackage{comment}
\usepackage{multirow, rotating}
\usepackage{tabularx, booktabs}
\usepackage{colortbl}
\usepackage{makecell}
\usepackage{longtable}
\usepackage{hyperref}
\usepackage[bottom]{footmisc}
\usepackage{pdflscape}
\usepackage{afterpage}
\usepackage{makecell}
\usepackage{xcolor}
\usepackage{listings}
\usepackage{titling}
\usepackage{colortbl}
\usepackage{comment}
\usepackage{appendix}
\usepackage{array}

\newcolumntype{L}[1]{>{\raggedright\arraybackslash}p{#1}}
\newcolumntype{C}[1]{>{\centering\arraybackslash}p{#1}}
\newcolumntype{R}[1]{>{\raggedleft\arraybackslash}p{#1}}

\title{Generating density nowcasts for U.S. GDP growth with deep learning: Bayes by Backprop and Monte Carlo dropout} 

\author{
Kristóf Németh\thanks{
Corresponding author. 
Faculty of Electrical Engineering and Informatics, 
Budapest University of Technology and Economics; 
E-mail: \href{mailto:manzotta@gmail.com}{manzotta@gmail.com}} \and 
Dániel Hadházi\thanks{
Department of Measurement and Information Systems, 
Budapest University of Technology and Economics; 
E-mail: \href{mailto:hadhazi@mit.bme.hu}{hadhazi@mit.bme.hu}}
}
\date{\today}

\begin{document}

\maketitle

\begin{abstract}
Recent results in the literature indicate that artificial neural networks (ANNs) can outperform the dynamic factor model (DFM) in terms of the accuracy of GDP nowcasts. 
Compared to the DFM, the performance advantage of these highly flexible, nonlinear estimators is particularly evident in periods of recessions and structural breaks. 
From the perspective of policy-makers, however, nowcasts are the most useful when they are conveyed with uncertainty attached to them. 
While the DFM and other classical time series approaches analytically derive the predictive (conditional) distribution for GDP growth, ANNs can only produce point nowcasts based on their default training procedure (backpropagation). 
To fill this gap, first in the literature, we adapt two different deep learning algorithms that enable ANNs to generate density nowcasts for U.S. GDP growth: 
Bayes by Backprop and Monte Carlo dropout. 
The accuracy of point nowcasts, defined as the mean of the empirical predictive distribution, is evaluated relative to a naive constant growth model for GDP and a benchmark DFM specification. 
Using a 1D CNN as the underlying ANN architecture, both algorithms outperform those benchmarks during the evaluation period (2012:Q1 -- 2022:Q4). 
Furthermore, both algorithms are able to dynamically adjust the location (mean), scale (variance), and shape (skew) of the empirical predictive distribution. 
The results indicate that both Bayes by Backprop and Monte Carlo dropout can effectively augment the scope and functionality of ANNs, rendering them a fully compatible and competitive alternative for classical time series approaches. 
\\

Journal of Economic Literature (JEL) codes: C32, C45, C53 

\end{abstract}

\section{Introduction}  \label{sec:intro}

Gross Domestic Product (GDP) is arguably the most important flow-type monetary measure of economic activity over a given period. 
GDP is often calculated on an annual basis, but in most advanced economies it is also measured on a quarterly frequency. 
Considering its relevance, the availability and reliability of these quarterly data are of great significance. 
However, due to the time required for data collection and statistical data processing, statistical offices are delayed in releasing GDP data for the current quarter. 
The first measurements are usually available  only a few months after the end of the current quarter, which makes it difficult to assess economic activity in real-time (i.e., synchronously). 
Therefore, many decision makers (such as central banks, budget offices, etc.) rely on the current quarter forecast, that is, the nowcast of GDP growth for more informed policy making. 
This underscores the need for accurate nowcasts. 

In nowcasting current quarter GDP growth, one typically tries to extract the informational content of higher frequency monthly data to track the real-time development of economic activity. 
Such an analysis faces the following three problems \citep{giannone2008nowcasting}: (i) Combining monthly predictor variables and a quarterly target variable. 
(ii) Handling a large number of potential regressors. 
(iii) Monthly data are released at different times (unsynchronized) within a quarter. 
This leads to the so-called \textit{``ragged edge''} problem, where some monthly features may have missing values, usually at the end of the sample period \citep{wallis1986forecasting}.

Capturing underlying relationships in macroeconomic data with a few latent components was first proposed by \textcite{rhodes1937construction}, who suggested replacing the Business Activity indicator of The Economist with the first principal component of the series used for its calculation. 
\textcite{stock1989new, stock2002forecasting}, \textcite{forni2000generalized}, \textcite{mariano2003new}, and \textcite{doz2012quasi} contributed to the revival and refinement of this approach. 
The use of the dynamic factor model (hereinafter, DFM) for GDP nowcasting was first formalized by \textcite{giannone2008nowcasting}. 
As proposed by \textcite{giannone2008nowcasting}, the DFM is a robust and flexible nowcasting model, able to accommodate for missing observations, jagged patterns of data, and mixed frequencies as well. 

A popular method for estimating unobserved components such as common factors consists of writing the model in state space form and using the Kalman filter to evaluate the model’s log likelihood. 
The Kalman filter provides an efficient approach for estimating nowcasting models for two reasons. 
First, it handles easily the missing values that arise when modelling jointly series aggregated at different frequencies and released asynchronously. 
Secondly, predictions can be decomposed into latent components, which notably can be used to capture secular changes in addition to common factors \citep{durbin2012time, harvey1990forecasting}. 
A comprehensive presentation of DFMs and state space methods for nowcasting is given by \textcite{banbura2013now}. 
As far as classical time series analysis goes, the DFM is still widely considered as one of the main nowcasting models. 

However, there are two important limitations of the model: 
(i) the almost always assumed linear structure, and (ii) the limited scalability of these models due to the computational challenges that are encountered when estimating factors models with more than a few dozens of variables. 
Non-Gaussian features can be introduced using importance sampling methods but these can be computationally intensive. 
Especially so, if we apply with the presence of multiple regressors. 
Actually, these two limitations together makes it difficult to effectively extend the model's capabilities. 
Alternatively, \textcite{creal2013generalized} and \textcite{harvey2013dynamic} derive a new class of filters relying on the score of the predictive log likelihood function which can arise from a wide range of families. 
Score driven models provide a general framework for introducing time-variation and latent states in any parameter of the predictive distribution, not only the location. 
\textcite{labonne2020capturing} uses common factors to exploit the cross-sectional relationships in location, scale, and shape parameters between GDP and a timely related series (monthly employment data). 
Based on a mixed-measurement score driven model and the asymmetric Student-t distribution, he extends the general idea behind dynamic factor models to conditional dispersion and asymmetry. 

As the recent results of \textcite{loermann2019nowcasting} and \textcite{hopp2021economic} show, ANNs can significantly outperform the dynamic factor model (DFM) in terms of accuracy of point nowcasts. 
These early results for ANNs are very promising and show great potential for these kind of applications. 
It is also worth noting, however, that significant performance gain over the DFM is achieved by readjusting the MLP's architecture along the rolling estimation window \citep{loermann2019nowcasting}. 
Unlike the recursive estimation of the dynamic factor model, not only the parameters but also the specification of the MLP is re-optimized in each step. 
This is done by an automated and algorithmic way with a combined filter - wrapper method described in \textcite{crone2010feature}, but as it stands now, this technique is only available for the MLP. 
\textcite{hopp2021economic} uses an LSTM, so it does not apply automated filters and wrappers, but the results reported are not for GDP growth. 

As accurate as they are, there is a major limitation of those ANN-generated GDP nowcasts, which can be seen from the perspective of the policy makers, who actually make use of them. 
That perspective is clearly recognized and summarized by the former Chair of the Federal Reserve, Alan Greenspan, as follows: 
``\textit{A central bank needs to consider not only the most likely future path for the economy, but also the distribution of possible outcomes about that path}'' \citep[37]{greenspan2004risk}. 
It would be hard to argue that nowcasts are most useful when they are conveyed with a measure of uncertainty attached to them. 
The DFM and the other classical time series approaches (e.g., MIDAS regression, score-driven models) provide distinct ways to generate not only point nowcasts, but also to attach an associated level of uncertainty to them. 
For example, the DFM, given its state space specification, uses the Kalman filter to predict the conditional mean (location) and standard deviation (scale or dispersion) of the measurement (target) variable in every time step. 
The predictive distribution of GDP growth, i.e. the density nowcast, is associated with that conditional (normal) distribution which is based on the information set available before measurement. 
By contrast, ANNs can only produce point nowcasts based on their default training procedure, i.e., backpropagation. 

In this paper we elevate this limitation by applying two different deep learning algorithms (DL algorithms): 
Bayes by Backprop and Monte Carlo dropout. 
These novel DL algorithms can effectively augment the scope and functionality of ANNs, enabling them to quantify the associated uncertainty of GDP point nowcasts. 
With ANNs, to estimate uncertainty requires generating the whole predictive distribution of the target variable. 
This is right away a very important difference compared to the DFM or other models of classical time series analysis. 
Those models rely on some kind of assumptions about the distribution of the target variable: 
As usual, our benchmark DFM specification, which can be seen in many applications, assumes quarterly GDP growth to follow a Gaussian distribution. 
In the case of the Gaussian, or other parsimoniously parameterized distributions (e.g., Student's t distribution) we could provide a summary measure of the scale of the predictive distribution. 
Namely, the standard deviation (variance). 
Since ANNs impose no prior assumptions on the distribution of the target variable, we need to generate the whole predictive distribution. 
First, we investigate the application of Bayesian neural networks. 

Bayesian neural networks (BNNs) are probabilistic models of their target variable ($\mathbf{y}$), where the predictive distribution of the target variable is defined as $P(\mathbf{y} | \mathbf{x}, \mathbf{w})$: 
Given an input $\mathbf{x} \in \mathbb{R}^{p}$, a neural network assigns a probability to each possible output $y \in \mathcal{Y}$, using the set of parameters or weights $\mathbf{w}$.
For classification, $\mathcal{Y}$ is a set of classes and $P(\mathbf{y} | \mathbf{x}, \mathbf{w})$ is a categorical distribution. 
For our regression-based predictive analysis, $\mathcal{Y}$ is $\mathbb{R}$ and $P(\mathbf{y} | \mathbf{x}, \mathbf{w})$ is continuous. 
With BNNs, instead of optimizing the trainable parameters (weights), we want to learn the posterior distribution of the weights $P(\mathbf{w} | \mathcal{D})$, where $\mathcal{D}$ stands for the set of training examples: $\mathcal{D} = \{ (\mathbf{x}_i, \mathbf{y}_i) \}$. 
After learning (estimating) the posterior distribution of the weights, we obtain the predictive distribution of the target variable through random sampling. 
This distribution answers predictive queries about unseen data by taking expectations: the predictive distribution of an unknown label $\mathbf{y}$ corresponding to a test data item $\mathbf{x}$, is given by $P(\mathbf{y} | \mathbf{x}) = E_{P_{(\mathbf{w} | D)}} \left[ P(\mathbf{y} | \mathbf{x}, \mathbf{w}) \right]$. 
Intuitively, each possible configuration of the weights, weighted according to the posterior distribution, makes a prediction about the unknown label given the test data item $\mathbf{x}$. 
Thus taking an expectation under the posterior distribution on weights is
equivalent to using an ensemble of an uncountably infinite number of neural networks. 
Unfortunately, the great promise of BNNs is usually undermined for neural networks of any practical size because we cannot learn the true posterior distribution of the weights from limited training data. 
Bayes by Backprop, proposed by \textcite{blundell2015weight} is an efficient, backpropagation-compatible algorithm that uses variational inference to approximate the theoretical posterior distribution of the weights. 
It is worth noting that, even the additional computational cost of Bayes by Backprop can be prohibitive for many applications. 
Without delving into the computational details, we only give the basic intuition here. 
We will describe Bayes by Backprop in details in Section \ref{sec:bayes_by_backprop}. 

Let us suppose that we already have a pre-trained ANN, trained with classical backpropagation, that generates point nowcasts that are accurate enough for our use case. 
In line with the aim of this study, the straightforward extension of this model would be to make it able to generate density nowcasts as well. 
Following the Bayesian approach, first we need to place a prior distribution upon the weights. 
Even in the most simple case, where we assume a normal distribution with a diagonal covariance matrix, we would have double the number of weights as in the original network. 
Considering the bias-variance trade-off, we shall expect worse generalization capabilities from the Bayesian model and so typically worse performance regarding the accuracy of point nowcasts. 
It is therefore not surprising that many empirical applications, trying to account for model uncertainty, use other methods of ensemble learning instead of Bayesian neural networks. 
Ensemble learning is a general machine learning technique that enhances accuracy and resilience in forecasting (nowcasting) by combining predictions from multiple models. 
It aims to mitigate errors or biases that may exist in individual models by leveraging the collective intelligence of the ensemble. 
If we think about ANNs, it looks promising to combine the results from different pre-trained models. 
It supposedly leads to better accuracy. 
If we refer to the combinations of these results as a specific sampling algorithm, we might end up with an empirical predictive distribution. 
There is a major difference between ensemble learning and BNNs though that we should emphasize. 
While different methods of ensemble learning can also generate a predictive distribution for the target variable, they do not learn it in a mathematically grounded way as Bayesian inference does.  

As we have seen, Bayes by Backprop generates the (approximate) predictive distribution of the target variable by learning the (approximate) posterior distribution of the weights. 
Monte Carlo dropout, the other DL algorithm adopted in the empirical analysis, generates the approximate predictive distribution via random sampling. 
As the name of the algorithm suggests, it presumes an underlying ANN that can be trained with dropout regularization, i.e., it must have dropout layers. 
Dropout regularization is a popular method used during the training phase to prevent overfitting in deep neural networks. 
During training, some layer outputs are randomly ignored or dropped according to the dropout rate ($p$).
Monte Carlo dropout, introduced by \textcite{gal2016dropout}, is a technique that extends the use of dropout to the inference phase of a neural network. 
It involves making multiple predictions for each input by applying different dropout masks, where each dropout mask ($\mathbf{z_i}$) is sampled from a multivariate Bernoulli distribution: $\mathbf{z_i \sim \mathbb{B}(1-p)}$. 
These dropout masks define different sets of neurons that are dropped out, so each prediction we make is generated by a (slightly) different network architecture. 
Then, the resulted predictive distribution can be used for generating point forecasts (e.g., by computing the mean) and for assessing the uncertainty related to the predictions (e.g., by computing the standard deviation). 
The number of predictions made for each input is a hyperparameter that can be adjusted to balance between accuracy and computational efficiency. 
One of the significant benefits of Monte Carlo dropout is that it provides a measure of uncertainty in the network's predictions. 
In addition, Monte Carlo dropout can improve the accuracy of the network's predictions. 
When making predictions, the average of the multiple predictions is taken to make a final prediction.
Instead of the heuristic scaling applied in inference time in classic dropout, MC Dropout samples from the same distribution of the networks in both inference and training time. Therefore, the average of the output of the sampled networks can produce more accurate point estimations in inference time. 
Since we train the shared weights of the sampled networks in parallel, Monte Carlo dropout is computationally more efficient than most other methods of ensemble learning. 

In the empirical analysis, we generate density nowcasts for the US GDP growth with a benchmark DFM and with two novel deep learning algorithms: 
With Bayes by backprop and Monte Carlo dropout. 
While the DFM derives those predictive densities analytically, i.e., in a closed form, the DL algorithms use sampling to generate empirical predictive densities for the target variable. 
The evaluation period ranges from 2012:Q1 to 2022:Q4, including periods both of stable economic growth and high economic turbulence. 
We perform a quasi real-time analysis, using real-time monthly vintages of the FRED-MD database for this period. 
Additionally, we distinguish three different information sets based on which nowcasts are conducted and evaluated. 
These information sets are related to the publication schedule of monthly vintages of the FRED-MD database. 
Specifically, they are defined by the intra-quarterly month in which the last vintage of FRED-MD was released. 
Together they form a 3-step nowcasting window for the empirical analysis. 
The accuracy of point nowcasts, defined as the mean of the empirical predictive distribution, is evaluated relative to a naive constant growth model for GDP and a benchmark DFM specification. 
Using a 1D CNN as the underlying ANN architecture, both algorithms outperform those benchmarks during the evaluation period. 
Furthermore, both algorithms are able to dynamically adjust the location (mean), scale (variance) and shape (skew) of the empirical predictive distribution. 
From the perspective of the policy-makers, those features can deliver additional insights into the current economic state. 
Density nowcasts generated by Monte Carlo dropout associate downturns (2020:Q2) with significant negative skew and recovery (2020:Q3) with significant positive skew. 
Similarly, Bayes by backprop produces a density nowcast for 2020:Q2 which shows significant negative skew at the second step of the nowcasting window (i.e., in May). 
Asymmetry of the predictive distributions can provide valuable additional information during critical periods, signaling the expected direction of the prediction error. 
The results indicate that both Bayes by Backprop and Monte Carlo dropout can effectively augment the scope and functionality of ANNs, rendering them a fully compatible and competitive alternative for classical time series approaches. 

The paper proceeds as follows: 
Section \ref{sec:data} describes the data behind the analysis and the feature selection strategies related to the different models and algorithms. 
Section \ref{sec:models_and_methods} describes our benchmark dynamic factor model and the two DL algorithms adopted in the empirical analysis: 
Bayes by Backprop, and Monte Carlo dropout. 
Section \ref{sec:empirical_analysis} presents the design of the nowcasting exercise. 
It defines the information sets based on which nowcasts are conducted and how density nowcasts are generated for the different models. 
Then, it presents and discusses the results of the empirical analysis. 
Finally, Section \ref{sec:conclusion} summarizes and concludes.

\section{Data and feature selection}    \label{sec:data}

\begin{figure}[H]
\begin{center}
\includegraphics[width = \textwidth]{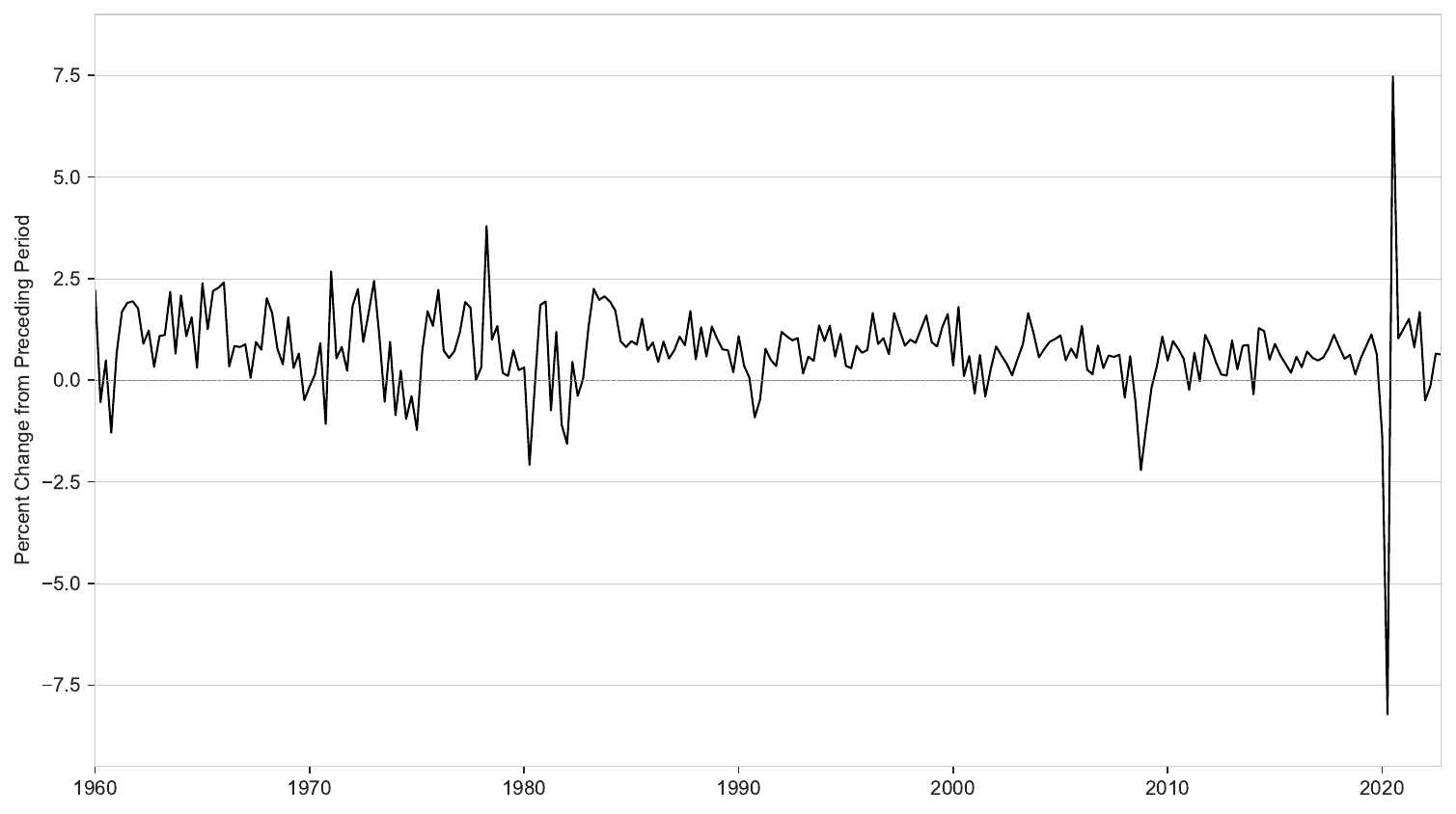}
\caption{The percent change (relative to the preceding period) of the US Gross Domestic Product. Source: Federal Reserve Economic Data (FRED).}
\label{fig:gdp_growth}
\end{center}
\end{figure}

The target variable of the empirical analysis is the quarterly GDP growth, that is the percent change (relative to the preceding period) of the US Gross Domestic Product. 
The series is measured on a quarterly frequency and is available from 1947:Q2 to 2022:Q3. 
Thus it contains 302 observations. 
Figure \ref{fig:gdp_growth} plots the time series for quarterly GDP growth. 

When the empirical analysis was closed, the last GDP measurement was released for 2022:Q4. 
In our empirical analysis, we try to extract the informational content of several monthly macroeconomic indicators (i.e., features) whose intra-quarterly development might be expressive of the current economic growth. 
The data for the input features comes from the FRED-MD database, the monthly database for Macroeconomic Research of the Federal Reserve Bank of St. Louis, which is described extensively in \textcite{mccracken2016fred}. 

FRED-MD is a large macroeconomic database that is designed for the empirical analysis of ``big data''. 
The database is publicly available and updated in real-time on a monthly basis. 
\footnote{The FRED-MD database is available for download under the following link: \url{https://research.stlouisfed.org/econ/mccracken/fred-databases/}}. 
It consists of 134 monthly time series (indicators or features) which are clustered into eight categories:
\begin{enumerate}
    \item Output and income (17 series)
    \item Labour market (32 series)
    \item Housing (10 series)
    \item Consumption, orders and inventories (14 series)
    \item Money and credit (14 series)
    \item Interest and exchange rates (22 series)
    \item Prices (21 series)
    \item Stock market (4 series)
\end{enumerate}

The complete list of the data and the series-related data transformations are described in detail in Appendix \ref{app:data}. 
Monthly indicator data are available from 1960:M1 to 2022:M12, so we have 756 observations for the potential regressors. 

For the empirical analysis, all monthly indicators (features) are detrended using the corresponding data transformations. 
This is important because all of our competitor models, except for the MLP, are designed to estimate time-invariant prediction rules, which is preferred for stationary time series. 
Consequently, when both the target variable and the regressors (features) are stationary, we can generally expect better predictive performance from these models compared to when modeling non-stationary series.  
Figure \ref{fig:rpi} below shows the level values of Real personal income (upper panel) and the result of the adequate data transformation (lower panel). 

\begin{figure}[H]
     \centering
     \begin{subfigure}[b]{0.9\textwidth}
         \centering
         \includegraphics[width=\textwidth]{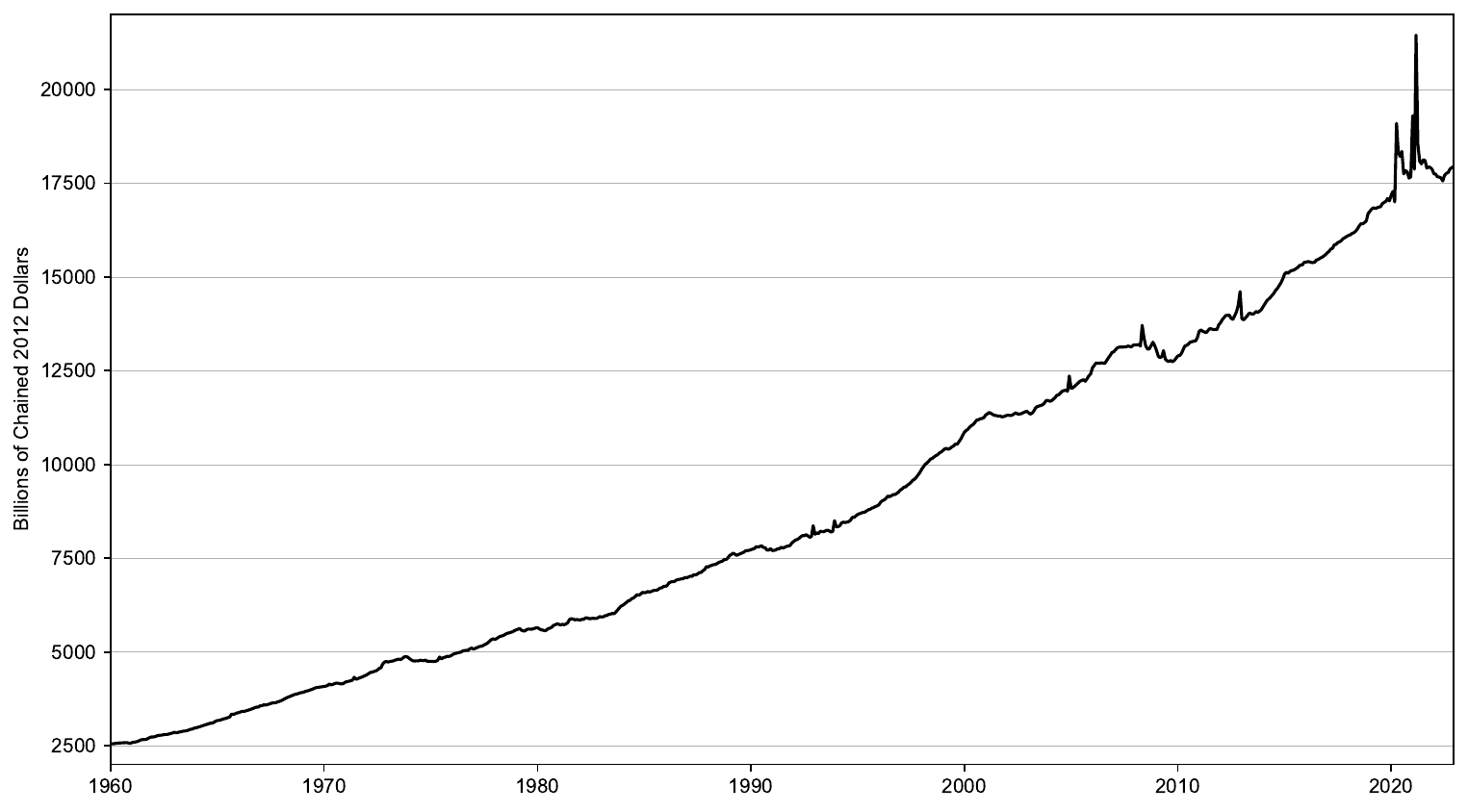}
         \caption{Real personal income (RPI): Original times series with trend.}
         \label{rpi_raw}
     \end{subfigure}
     \hfill
     \begin{subfigure}[b]{0.9\textwidth}
         \centering
         \includegraphics[width=\textwidth]{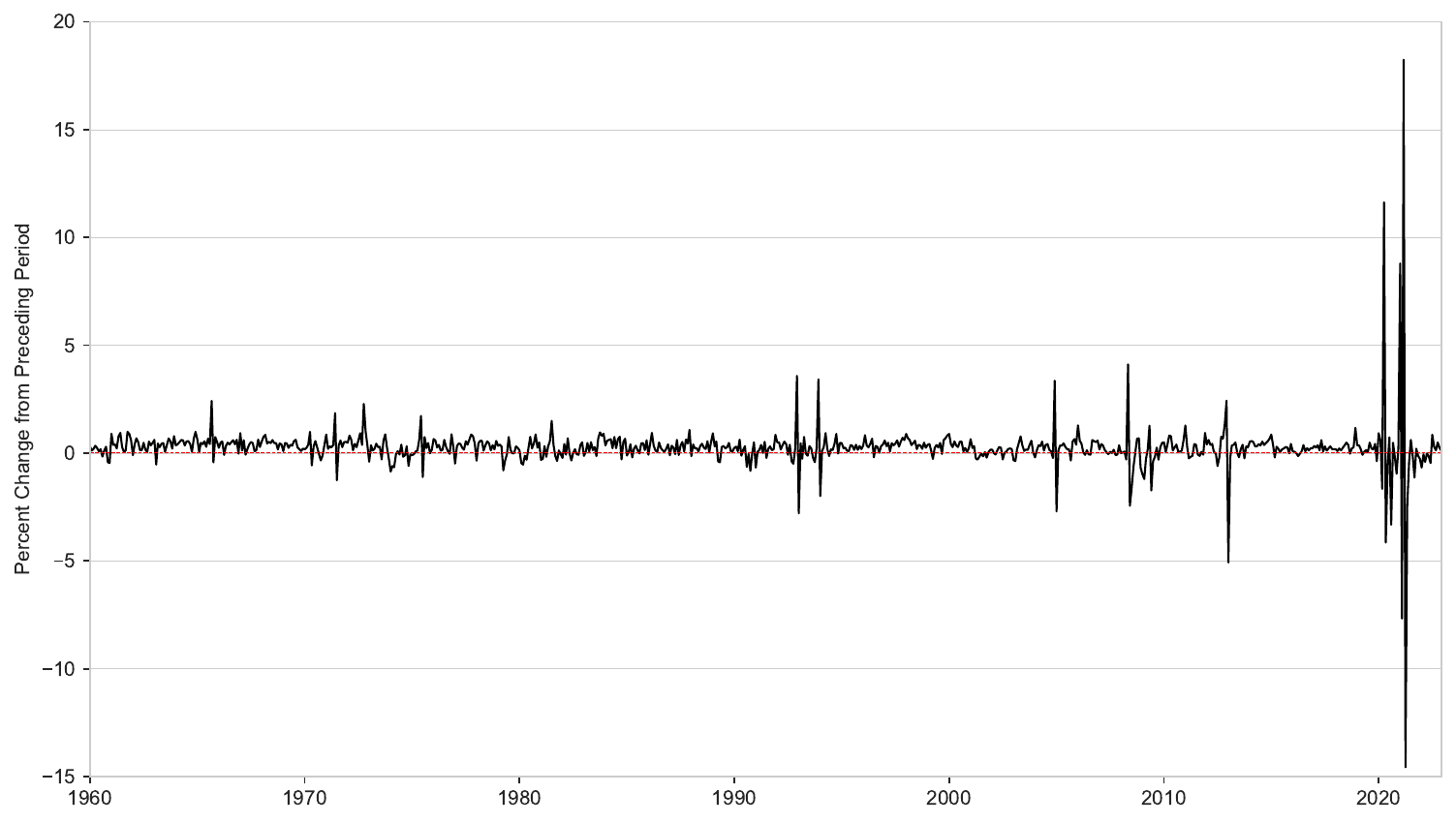}
         \caption{Real personal income (RPI): Log-differenced, detrended times series.}
         \label{rpi_transformed}
     \end{subfigure}     
        \caption{Real personal income (RPI). Source: FRED-MD.}
        \label{fig:rpi}
\end{figure}

At the end of this chapter, we should yet introduce our restricted dataset that has been used in the dynamic factor analysis. 
Unlike those neural network architectures investigated in our thesis, the dynamic factor model's computation time scales exponentially with the number of features \citep{hopp2021economic}. 
As we will see in Chapter 4, feature selection is also separated from the actual predictive step in the case of DFM. 
These two characteristics of DFM make feature selection a much more complicated task than is usually experienced in the case of neural networks.
Based on these considerations, we used the following four macroeconomic variables during the course of the dynamic factor analysis. 

\begin{itemize}
\item[-] Industrial production: Manufacturing (IPMANSICS)

\item[-] Real aggregate income (excluding transfer payments) (W875RX1)

\item[-] Manufacturing and trade sales (CMRMTSPL)

\item[-] Employees on non-farm payrolls (PAYEMS)
\end{itemize}

\begin{figure}[H]
        \centering
        \begin{subfigure}[h]{0.475\textwidth}
            \centering
            \includegraphics[width=\textwidth]{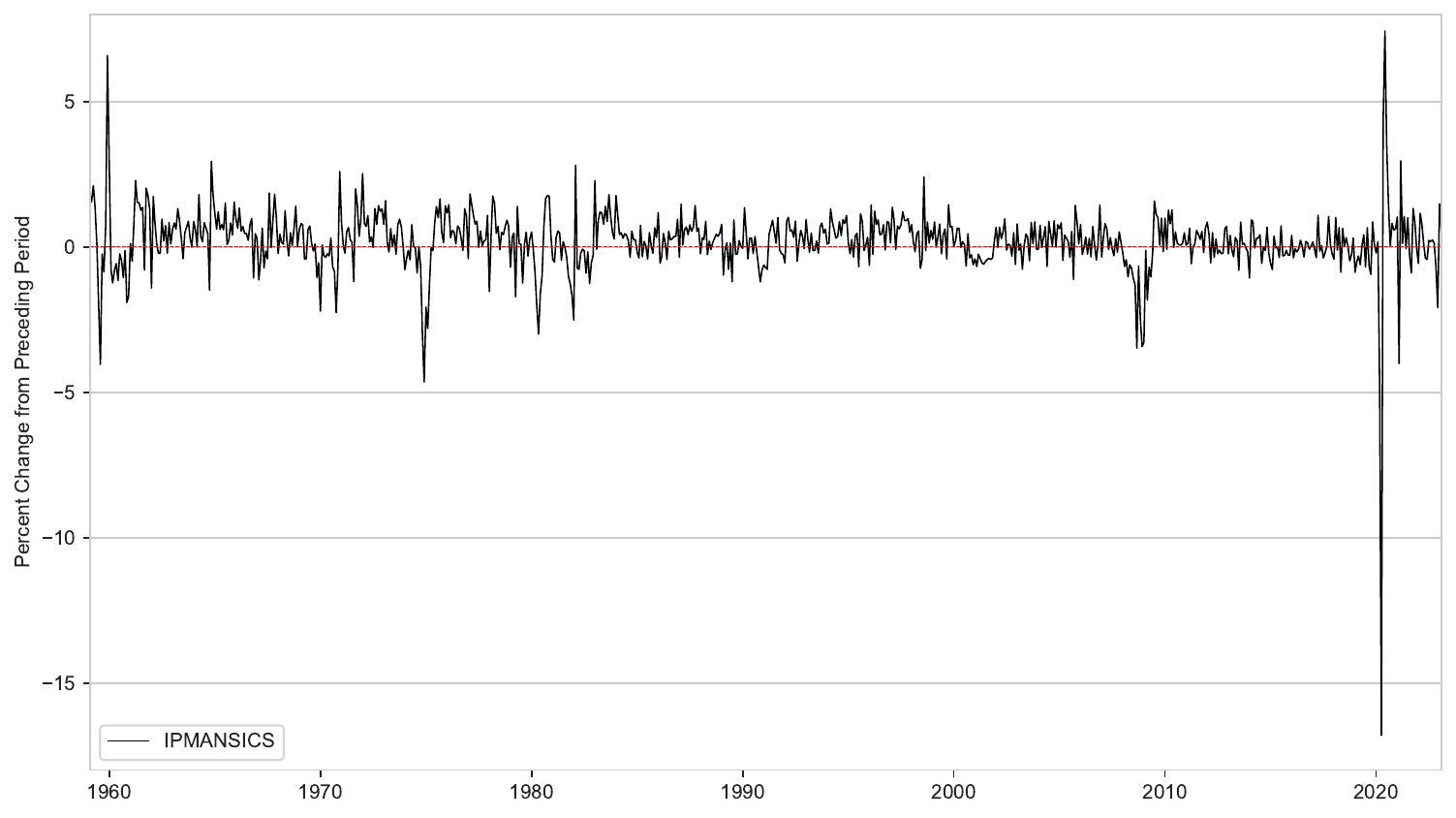}
            \caption[]%
            {{\small IP: Manufacturing (SIC)}}    
            \label{fig:data_ipmansics}
        \end{subfigure}
        \hfill  
        \begin{subfigure}[h]{0.475\textwidth}  
            \centering 
            \includegraphics[width=\textwidth]{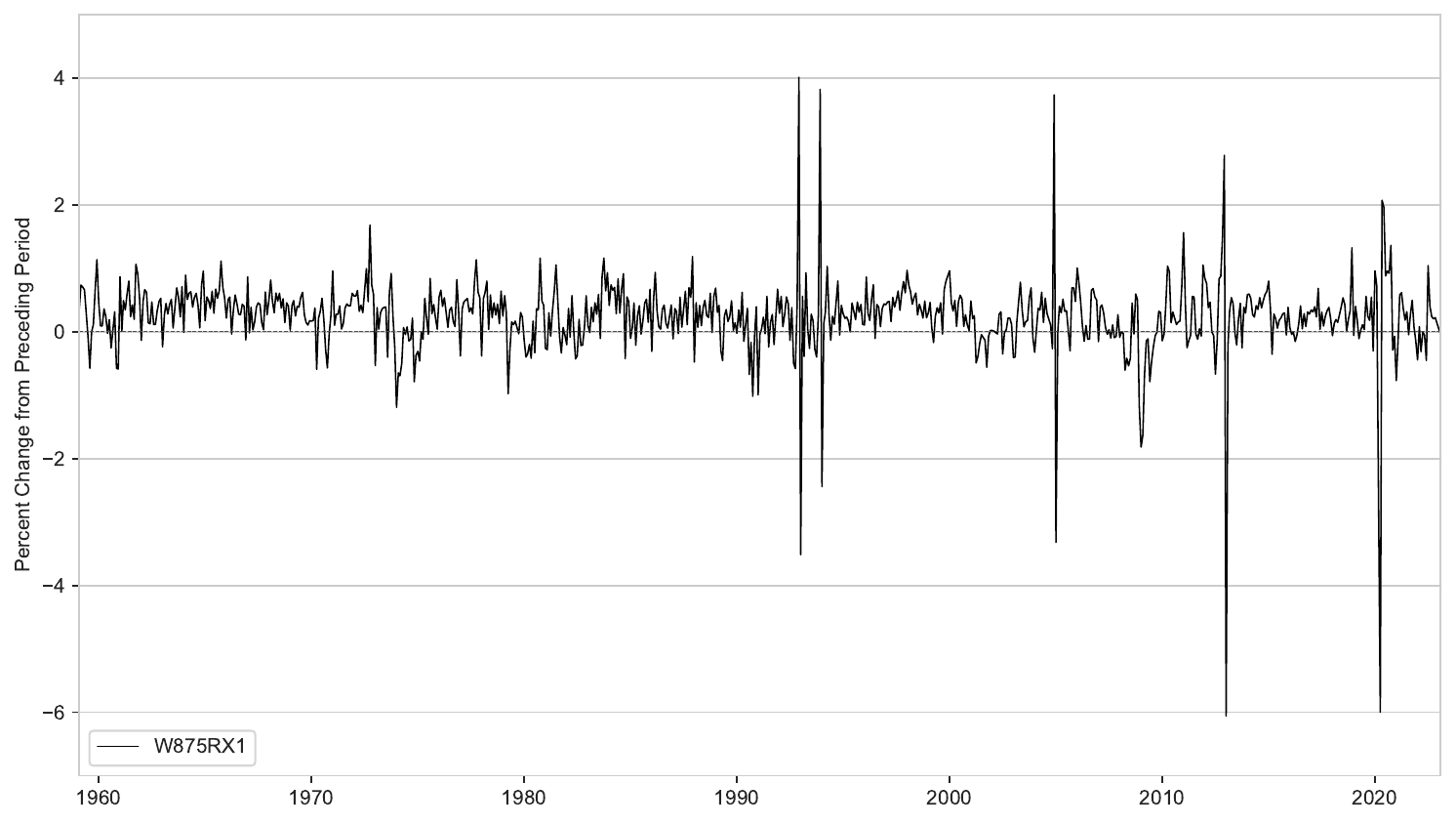}
            \caption[]%
            {{\small Real personal income ex. transfer receipts}}    
            \label{fig:data_w875rx1}
        \end{subfigure}
        \vskip\baselineskip
        \begin{subfigure}[h]{0.475\textwidth}   
            \centering 
            \includegraphics[width=\textwidth]{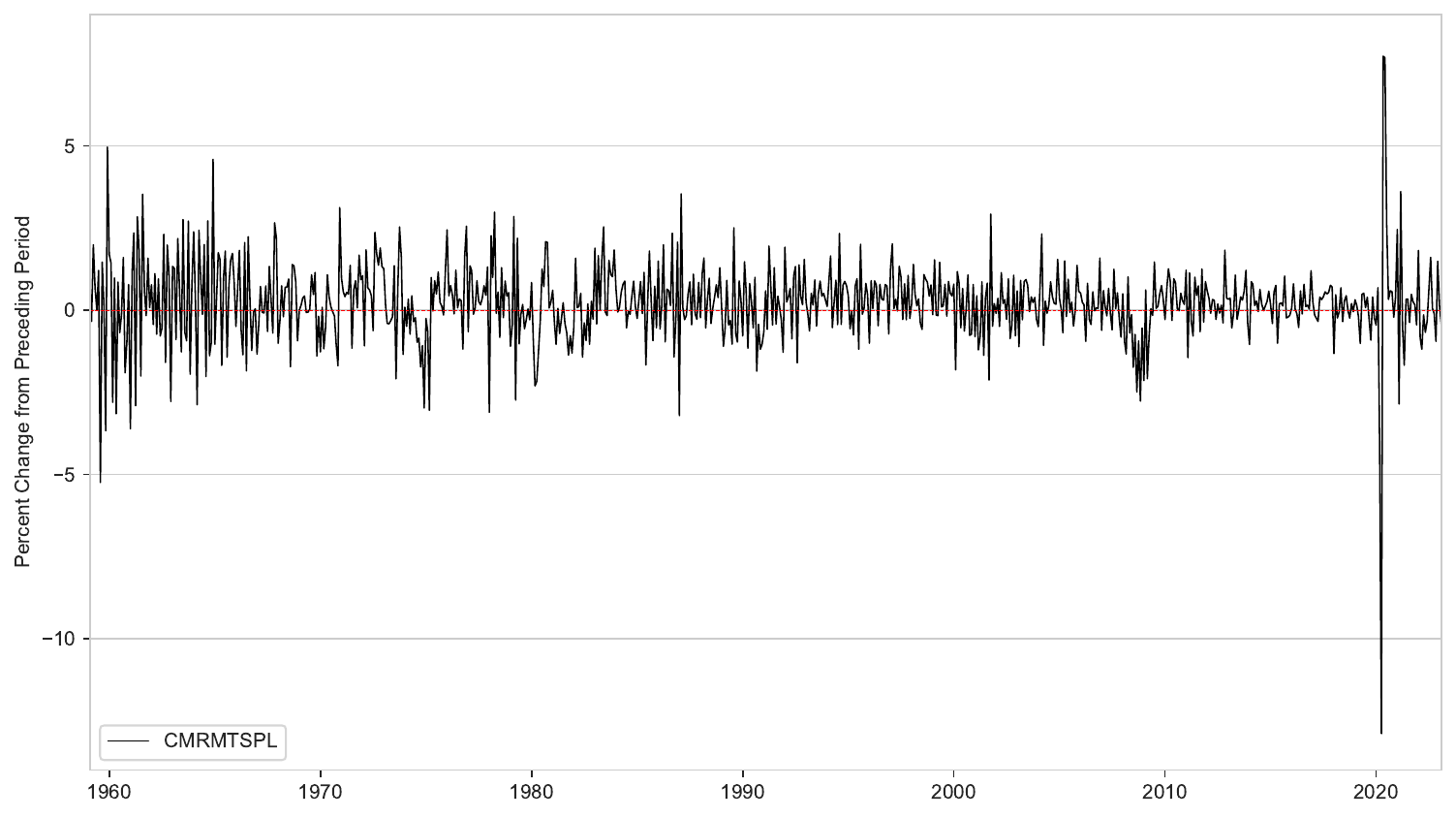}
            \caption[]%
            {{\small Real Manu. and Trade Industries Sales}}    
            \label{fig:data_cmrmrspl}
        \end{subfigure}
        \hfill
        \begin{subfigure}[h]{0.475\textwidth}   
            \centering 
            \includegraphics[width=\textwidth]{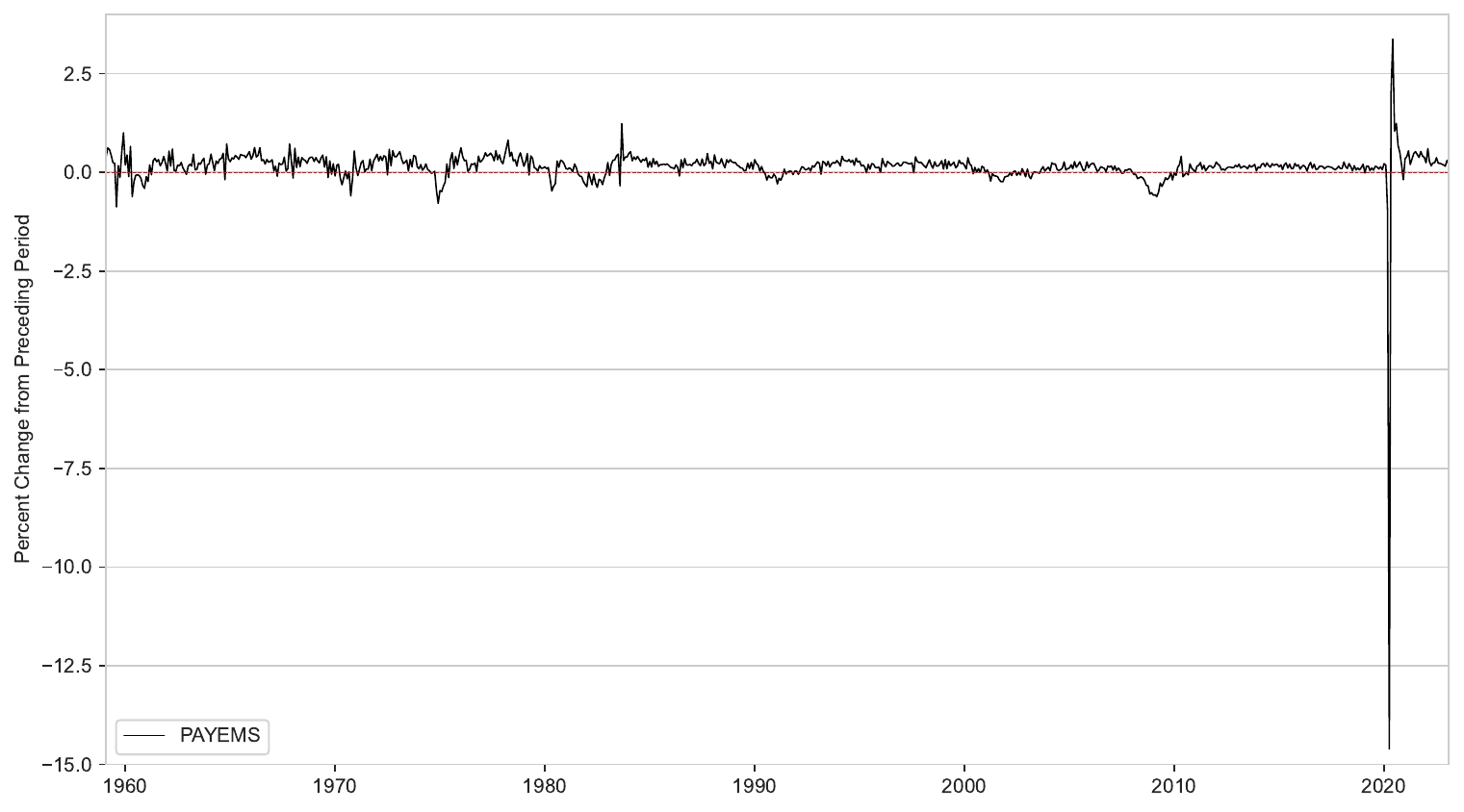}
            \caption[]%
            {{\small All Employees: Total nonfarm}}    
            \label{fig:data_payems}
        \end{subfigure}
        \caption[ The average and standard deviation of critical parameters ]
        {\small Time series used in the course of dynamic factors analysis. Source: Own calculations based on FRED-MD.} 
        \label{fig:dfm_features}
    \end{figure}

These series are proposed by the documentation of the \textit{statsmodels} Python module \citep{statsmodels}\footnote{Downloaded from: \url{https://www.statsmodels.org/dev/examples/notebooks/generated/statespace_dfm_coincident.html}}.  
All of the variables are also part of the FRED-MD database. 
Figure \ref{fig:dfm_features} shows our restricted dataset used exclusively during the estimation of the dynamic factor model.  

For the training of the different ANNs, we use all the monthly indicator series included in the FRED-MD database. 
At the same time, we introduce a bottleneck (encoder) layer, as the first layer in each network, to reduce the dimensionality of the feature space. 
The resulting low-dimensional embedding (state) vector contains a linear combination of the original variables, so in this respect, it performs a transformation on the data similar to principal component analysis (PCA). 
In the case of PCA, however, the weights of the resulting linear combination are determined to minimize the reconstruction error. 
Concretely, the weights used in constructing the $i$-th principal component are the elements of the eigenvector belonging to the $i$-th largest eigenvalue of the covariance matrix. 
Compared to this, in the solution we propose, the weights of the linear combination are trained in the same integrated training process.
Thus they are determined to minimize the prediction error. 
Considering the relatively small number of available training samples, this type of dimension reduction should significantly help each ANN's generalization capability. 
Besides dimension reduction, we also apply regularization to the weight vector for the bottleneck layer ($\mathbf{w}_b$). 
Equation (\ref{eq:6}) describes the cost function augmented with the an L1 (Lasso) regularization term \citep{jiang2016variable}: 
\begin{align}	\label{eq:6}
C(\mathbf{w}) = \frac{1}{N} \sum_{i=1}^{N} L \left( y_{i}, f(\mathbf{w}, \mathbf{x_i}) \right) + \lambda \lVert \mathbf{w}_b \rVert_{1} 
\end{align}
where $L()$ is the loss (criterion) function selected for training, $\lVert \mathbf{w}_b \rVert_{1}$ denotes the L1 norm of $\mathbf{w}_b$ and $\lambda$ is a hyperparameter which determines how severe the penalty is.\footnote{We will specify the loss function later, during the description of the training process. } 
Intuitively, a higher value of $\lambda$ induces a bigger penalty to the L1 norm thus it leads to a more sparse weight vector. 
As Equation (\ref{eq:6}) suggests, we do not use regularization in any other layer of of the different networks. 
This L1 (Lasso) regularization leads to a sparse weight vector in the bottleneck (encoder) layer and ultimately results in feature selection \citep{jiang2016variable}. 
Figure \ref{fig:weights} below shows the histograms of the estimated parameters (weights and biases) for the bottleneck layer ($\mathbf{w}_b^{*}$) without regularization (\ref{fig:weights_default}) and with L1 regularization(\ref{fig:weights_l1}). 

\begin{figure}[H]
\centering
   \begin{subfigure}[t]{0.45\linewidth}
   \centering
   \includegraphics[width=\linewidth]{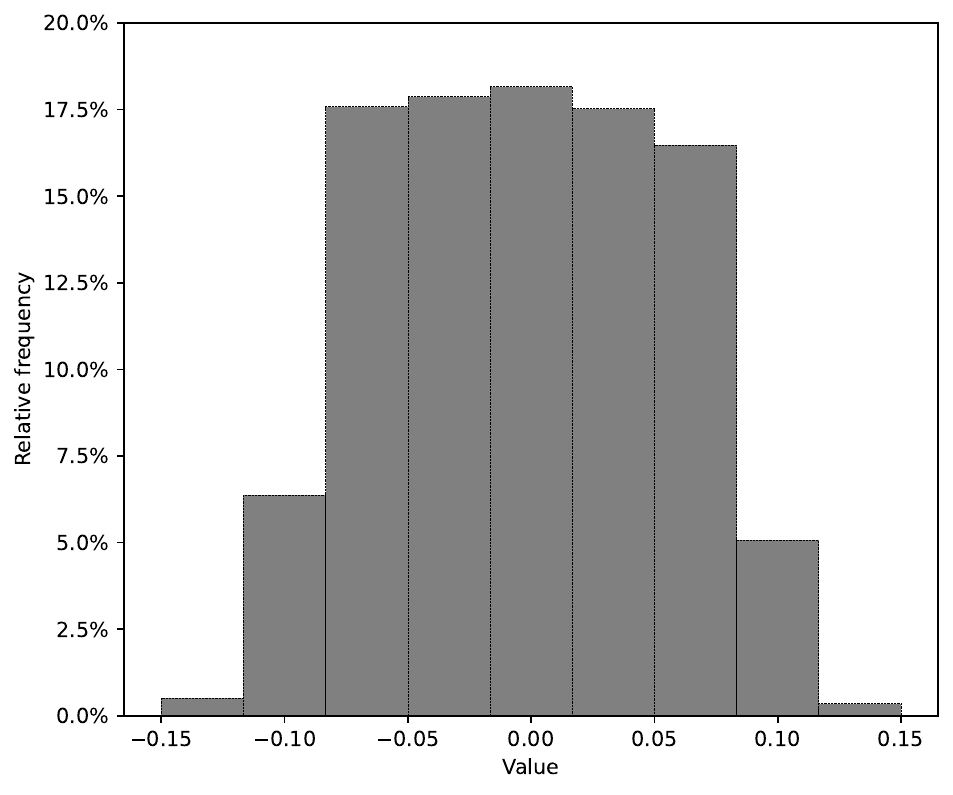}
   \caption{Histogram of $\mathbf{w}_b^{*}$ without regularization.}
   \label{fig:weights_default} 
\end{subfigure}
\hfill
\begin{subfigure}[t]{0.45\linewidth}
   \centering
   \includegraphics[width=\linewidth]{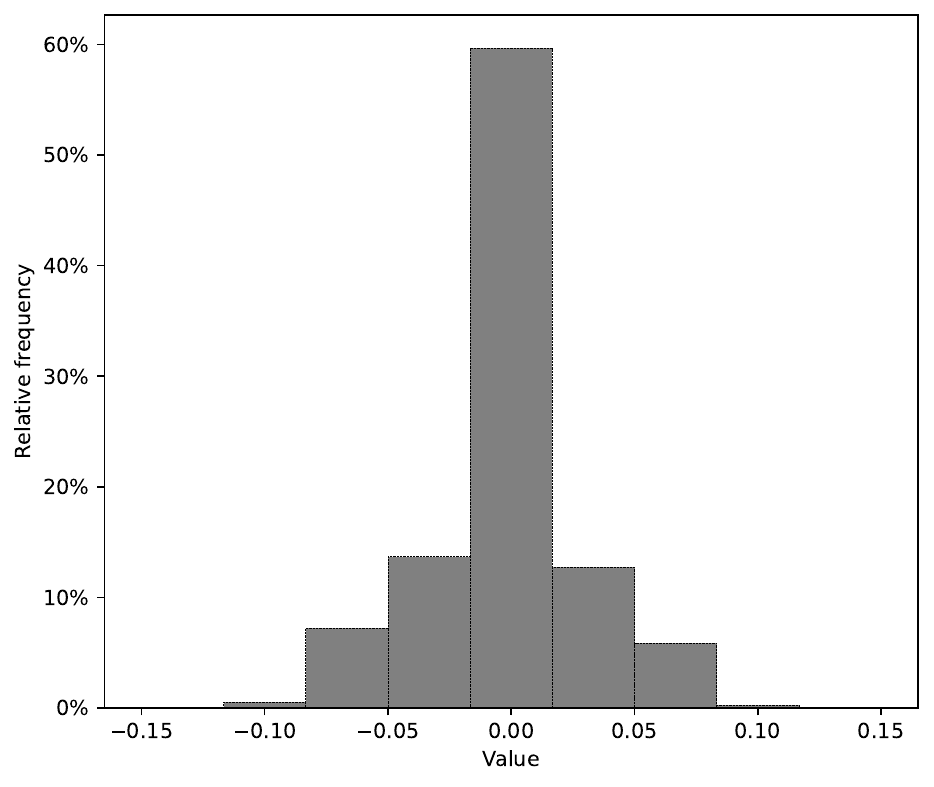}
   \caption{Histogram of $\mathbf{w}_b^{*}$ with L1 regularization.}
   \label{fig:weights_l1}
\end{subfigure}
\centering
\caption{Empirical distribution of the estimated parameters (weights and biases) for the bottleneck (encoder) layer in the 1D CNN. Source: Own editing based on FRED-MD.}
\label{fig:weights}
\end{figure}

\section{Models and methods}    \label{sec:models_and_methods}

This section presents the models and algorithms adopted in the empirical analysis. 
First, we describe the dynamic factor model, one of the state-of-the-art nowcasting models regarding classical time series approaches. 
Then, we describe those two deep learning algorithms based on which density nowcasts are generated for U.S. GDP growth: 
Namely, Bayes by Backprop and Monte Carlo dropout.

\subsection{Dynamic factor model}   \label{sec:dfm}

The dynamic factor model (hereafter, DFM) is a state space model, the main characteristic of which is that the observation vector is much larger in dimension than the vector of the unobserved (latent) common factors.  
The general specification of the DFM can be written as follows: 
\begin{align*}
x_t &= \Lambda f_t + B d_t + u_t \\
u_t &= C_1 u_{t-1} + \dots + C_q u_{t-q} + \varepsilon_t \qquad \varepsilon_t \sim N(0,\Sigma) \\
f_t &= A_1 f_{t-1} + \dots + A_p f_{t-p} + \eta_t \qquad \eta_t \sim N(0,I) 
\end{align*}
where $x_t$ are observed indicator data, $f_t$ are the unobserved common factors (evolving as a vector autoregression), $d_t$  are (optional) exogenous variables, and $u_t$ is the error, or ``idiosyncratic'' process ($u_t$ is also optionally allowed to be autocorrelated). 
The $\Lambda$ matrix is also referred here to as the matrix of \textit{factor loadings}. 
The variance of the factor error term is set to the identity matrix to help the identification of the unobserved factors. 
The size of the observation vector is typically large while the dimension of the state vector is small, so that we have $p >> m$ \citep{durbin2012time}. 

This model can be cast into state space form, then the unobserved factors are estimated via the Kalman filter. 
The likelihood is evaluated as a byproduct of the filtering recursions, and maximum likelihood estimation used to estimate the parameters. 
The (log)likelihood function of the unknown parameters can be formulated as follows \citep{harvey1990forecasting}:

\begin{align}   \label{eq:ssm_loglikelihood}
\log{L} = -\frac{NT}{2} \log{2\pi} - \frac{1}{2} \sum_{t=1}^{T} \log{|{F_t}|} - \frac{1}{2} \sum_{t=1}^{T} {{v}_{t}^{'} F_{t}^{-1} {v}_t}
\end{align} 
where $N$ denotes the dimension (width) of the observation vector $y_t$. 
The Kalman filter pre-fit residual $v_t = y_t - E[y_t | Y_{t-1} ]$ and its variance $F_t = Var(v_t)$ are the one-step ahead forecast error and the one-step ahead forecast error variance of $y_t$ given $Y_{t-1}$. 
Accordingly, $Y_{t-1}$ is the information set available at time $t-1$ or, more precisely, right before the measurement for $y_t$ gets realized: $Y_{t-1} = \{ y_{t-1}, y_{t-2}, \dots, y_{1}; \ x_{t-1}, x_{t-2}, \dots x_{1}\}$. 
The forecast errors $v_1, \dots, v_T$ are sometimes called innovations because they represent the new part of $y_t$ that cannot be predicted from the past for $t = 1, \dots, T$ \citep{durbin2012time}. 
The quantities ${v}_t$ and $F_t$ are calculated routinely by the Kalman filter, thus the log-likelihood is easily computed from the Kalman filter output. 
We assume that $F_t$ is non-singular for $t = 1, \dots, T$. 
If this condition is not satisfied initially it is usually possible to redefine the model so that it is satisfied. 
The representation (\ref{eq:ssm_loglikelihood}) of the log-likelihood was first given by \textcite{schweppe1965evaluation}. 
\textcite{harvey1990forecasting} refers to it as the \textit{prediction error decomposition}. 

\begin{figure}[H]
\begin{center}
\includegraphics[width = \textwidth]{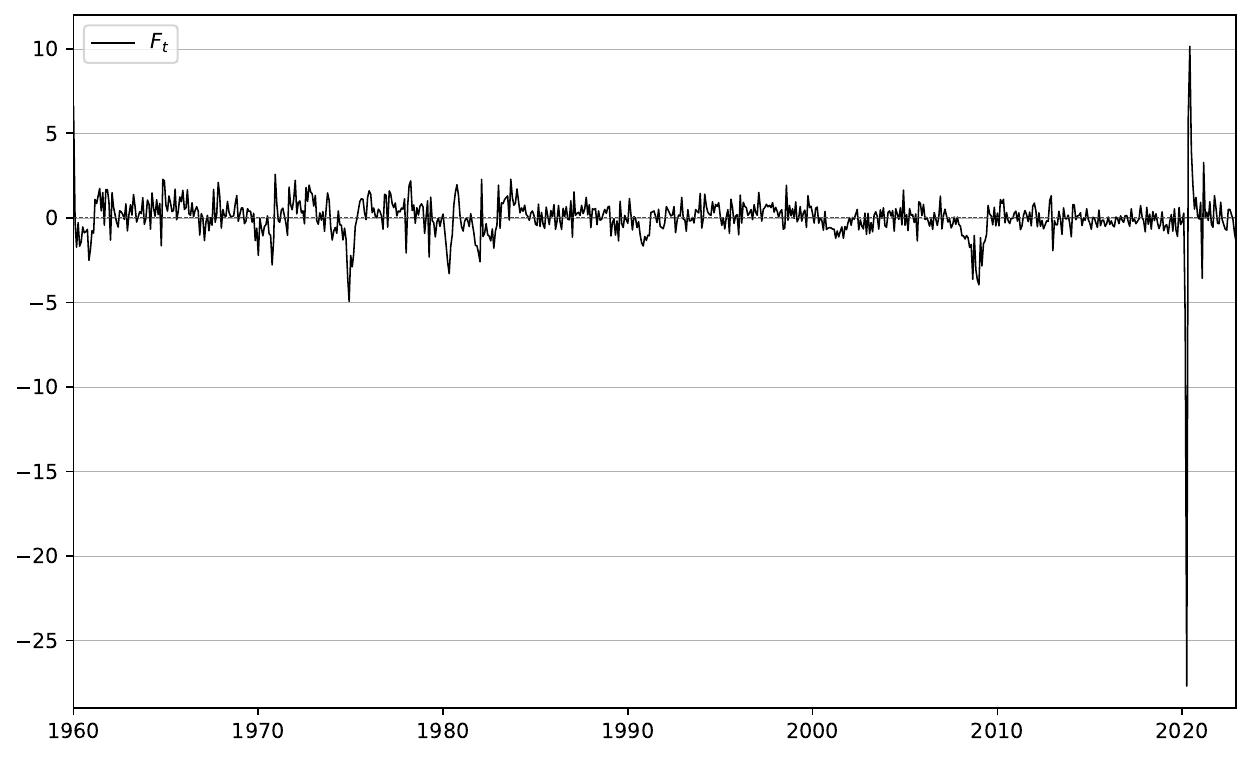}
\caption{Estimation of the DFM specification given by (\ref{eq:dfm_measurement})--(\ref{eq:dfm_factor}): 
Full-sample filtered estimates of one unobserved common factor ($f_t$). 
Source: Own calculation based on FRED-MD.}
\label{fig:dfm_result}
\end{center}
\end{figure}

The specific dynamic factor model in our empirical analysis has only $1$ unobserved common component which follows an AR(1) process. 
Similarly, we model the idiosyncratic component for each series as an AR(1) process, that are assumed to be cross-sectionally independent, so that $\Sigma$ is a diagonal matrix. 
 \begin{align}
x_{j, t} &= \Lambda f_{t} + u_{j, t}  
\label{eq:dfm_measurement} \\
u_{j,t} &= c_{1, j} u_{j, t-1} + \epsilon_t \qquad \epsilon_t \sim N(0, \sigma^{2}_{j}) 
\label{eq:dfm_idiosyncratic} \\
f_t &= a_{1} f_{t-1} + a_{2} f_{t-2} + \eta_t \qquad \eta_t \sim N(0,\sigma^{2}_{f})  
\label{eq:dfm_factor}
\end{align}
In Figure \ref{fig:dfm_result}, we can see the result of the dynamic factor analysis, that is, the estimation of the state space model described by equations (\ref{eq:dfm_measurement})--(\ref{eq:dfm_factor}). 
Values of the extracted factor (i.e., factor scores) refer to the filtered estimates of the unobserved state variable, $f_t$. 
Here we plot the result of a full-sample estimation to facilitate the comparison between Figure \ref{fig:gdp_growth} in Section \ref{sec:data} and Figure \ref{fig:dfm_result} below. 
Seeing those Figures together indicates how relevant the extracted factor is in terms of economic growth. 

\subsubsection*{Two-step nowcasting procedure based on DFM}

We use the DFM in a two-step nowcasting procedure: 
The common factor is estimated first based on the state space model defined by (\ref{eq:dfm_measurement})--(\ref{eq:dfm_factor}), then nowcasting is carried out using another separate state space model, basically functioning as a bridge regression.\footnote{It is worth noting that the state-space approach allows in general for joint estimation of the factors and nowcasting GDP \citep{banbura2007look, mariano2003new, marcellino2010factor}.} 
The specification of this latter state space model is also linear, both in its state and measurement equation, and assumes normally distributed error terms. 
While GDP growth is measured only on a quarterly frequency, i.e., in every third month, the model itself is cast at monthly frequency. 
Accordingly, the specification of the state dynamics is given by equation (\ref{eq:ssm_state}):
\begin{align}
y_{t}^{m} = c + \varphi f_t + \varepsilon_t, \qquad \varepsilon_t \sim N(0, \sigma_{\varepsilon}^2), 
\label{eq:ssm_state}
\end{align}
where in month $t$, $y_{t}^{m}$ denotes the value of the unobserved (\textit{latent}) monthly GDP growth, and $f_t$ stands for the value of the previously extracted dynamic factor -- summarizing the information content of monthly indicators. 
We should emphasize that $f_t$ is used as an exogenous (input) variable in equation (\ref{eq:ssm_state}). 
Then the corresponding measurement equation formulates a temporal aggregation rule which connects the unobserved monthly growth to our actual target variable, i.e., quarterly GDP growth: 

\begin{align}
y_{t}^{q} = \frac{1}{3} \left( y_{t}^{m} + 2 y_{t-1}^{m} + 3 y_{t-2}^{m} + 2 y_{t-3}^{m} + y_{t-4}^{m} \right), 
\label{eq:ssm_measurement} 
\end{align}
where in month $t$, $y_{t}^{q}$ denotes the value of the quarterly GDP growth. 
The temporal aggregation described by (\ref{eq:ssm_measurement}) is based on a very close approximation between monthly and quarterly growth rates which was first proposed by \textcite{mariano2003new}. 
The main intuition behind (\ref{eq:ssm_measurement}) is based on the fact that GDP is a flow-type variable.
Consequently, the GDP of a given quarter must be equal to the sum of the corresponding monthly GDP values:
\begin{align*}
Y_{t}^{q} = Y_{t}^{m} + Y_{t-1}^{m} + Y_{t-2}^{m},
\end{align*}

We should emphasize that our model is cast at monthly frequency, so our measurement variable has valid observations in only every third time step. 
However, this is not an issue because the Kalman filter accommodates missing data by not updating filtered state estimates corresponding to missing observations. 
In other words, suppose there is a missing observation at period $t$. 
Then, the state forecast for period $t$ based on the previous $t-1$ observations and filtered state for period $t$ are equivalent. 
For simplicity, we remove the upper index of $y^{q}_{t}$, so we denote quarterly GDP growth in month $t$ with $y_{t}$ in the remainder of the paper. 
In state space models, the a priori prediction for the target (measurement) variable is given by its one-step-ahead conditional expected value: 
\begin{align}
\hat{\mu}_{t|t-1} = E \left[ y_{t} | Y_{t-1} \right], 
\label{eq:dfm_mean_definition}
\end{align}
where $Y_{t-1}$ is the information set containing previous measurements for the target variable, and all the data for exogenous input variables, that is available before $y_t$ is observed. 
If we define $Y_{t-1} = \{ y_{t-1}, y_{t-2}, \dots, y_{1}; \ f_{t}, f_{t-1}, f_{t-2}, \dots f_{1}\}$ then we implicitly assume that exogenous input data ($f_{t}$) are accessible before the measurement for $y_{t}$ gets realized. 
Based on equations (\ref{eq:ssm_state}) and (\ref{eq:ssm_measurement}), we can define $\hat{\mu}_{t|t-1}$ as follows: 
\begin{align}
\hat{\mu}_{t | t-1} = 3c + \frac{1}{3} \varphi \left( f_{t} + 2 f_{t-1} + 3 f_{t-2} + 2 f_{t-3} + f_{t-4} \right). 
\label{eq:dfm_mean_v1}
\end{align}
From the perspective of the empirical analysis, we should remember that nowcasts are specific current-quarter forecasts, where the information set contains regressor observations also from the current quarter. 
We also know that GDP growth is measured on a quarterly frequency, so we have valid (non-missing) target measurements at every third month. 
If $q$ denotes quarters and $t$ still stands for months, then valid target measurements are associated with time indices: $t = 3q = 3, 6, 9, 12, \dots$. 
Accordingly, we define $m$ as that specific intra-quarterly month until then regressor data are contained in the information set. 
Based on these considerations, we can reformulate equation (\ref{eq:dfm_mean_v1}) as follows: 
\begin{align}
\hat{\mu}_{3q | m} = 3c + \frac{1}{3} \varphi \left( f_{3q | m} + 2 f_{3q-1 | m} + 3 f_{3q-2 | m} + 2 f_{3q-3 | m} + f_{3q-4 | m} \right). 
\label{eq:dfm_mean}
\end{align}
The role of the information set available in month $m$ ($Y_m$) can be revealed if we also consider the first state space model defined by (\ref{eq:dfm_measurement})--(\ref{eq:dfm_factor}). 
Based on that benchmark DFM specification, we extract (monthly) common factors from several monthly indicators, which are then taken as exogenous input variables in the second state space model, explicitly used for nowcasting. 
In this DFM-based two-step estimation procedure the information set determines how much monthly indicator data can be used to estimate the common factors. 
For example, if $m = 3(q-1) + 2$ then $x_t$ in equation (\ref{eq:dfm_measurement}) can have valid regressor observations until the second month of the current quarter ($q$). 
Similarly to equation (\ref{eq:dfm_mean_definition}), let us denote the one-step-ahead conditional variance of $y_t$ with $\hat{\sigma}^{2}_{t|t-1}$: 
\begin{align}
\hat{\sigma}^{2}_{t|t-1} = Var(y_{t} | Y_{t-1}). 
\label{eq:dfm_variance_definition}
\end{align}
If we consider a target observation for a given quarter, then conditioning on the information set available in intra-quarterly month $m$, we can expand the definition as follows: 
\begin{align}
\hat{\sigma}^{2}_{3q | m} = \frac{19}{9} \left( \varphi^{2} Var(f_{3q | m}) \right) + \frac{19}{9} \sigma^{2}_{\varepsilon}. 
\label{eq:dfm_variance}
\end{align}

Here it is important to note that while the extracted (estimated) series of the common factor, $f_{t}$, changes over time, its variance is constant. 
This follows from the specification of equation (\ref{eq:dfm_factor}), that describes a time-invariant state equation. 
So our benchmark DFM specification leads to a predictive model for $y_t$ with conditional mean and constant volatility. 
So our benchmark DFM specification leads to a conditional mean and variance (volatility) model for $y_t$, where the latter assumed to be constant. 

The dynamic factor model described above has long been considered the state-of-the-art statistical modeling framework for economic nowcasting. 
This modeling framework compresses information from a large dataset into a few underlying factors, while at the same time applying Kalman-filtering techniques to fill up the missing data of the ragged edge within the dataframe. 
Although this approach provides a unified framework incorporating dynamic imputation (model-based interpolation) and nowcasting, the central predictor equations, (\ref{eq:ssm_state}) and (\ref{eq:ssm_measurement}) remain linear, potentially restricting the model from generalizing to non-linear patterns \citep{bartholomew2011latent}.

\subsection{Estimating uncertainty with ANNs}

Deep learning has gained tremendous attention in applied machine learning. 
However, ANNs are inherently designed to generate point estimates (point nowcasts), so they cannot capture model uncertainty based on their default training procedure, i.e., backpropagation. 
In the following, we present two different deep learning algorithms that can elevate this limitation: 
Namely, Bayes by Backprop and Monte Carlo dropout. 
It is worth noting that these algorithms are not stand-alone in the sense that both of them use an underlying ANN and backpropagation to generate empirical predictive distributions for the target variable, i.e., density nowcasts. 
We use a one-dimensional convolutional neural network (1D CNN) as the underlying ANN architecture for both of these algorithms.\footnote{In the context of time series analysis, 1D CNNs are often referred to as time-delay neural networks (TDNNs).}

\subsubsection{Bayes by Backprop}   \label{sec:bayes_by_backprop}

Bayes by Backprop, introduced by \textcite{blundell2015weight}, is a principled, efficient, backpropagation-compatible algorithm for learning a probability distribution on the weights $P(\mathbf{w} | \mathcal{D})$ of a neural network. 
The main idea of the algorithm is that it approximates $P(\mathbf{w} | \mathcal{D})$ with a so-called \textit{variational} distribution $q(\mathbf{w} | \theta)$ which can be learnt during the training process. 
Then, the predictive distribution for the target variable is obtained by the repeated sampling from the approximate posterior $q(\mathbf{w} | \theta^{*})$. 

Before \textcite{blundell2015weight}, \textcite{hinton1993keeping} and  \textcite{graves2011practical} already suggested to use variational rather than Monte Carlo methods to find the approximate Bayesian posterior distribution based on the speed of computation.  
Variational inference basically means that we define a simplified approximate distribution $q$ with its variational parameters $\theta$ that shall be as similar as possible to the underlying true posterior distribution $P$ that is intractable:
\begin{align}
q(\mathbf{w} | \theta) &\approx P(\mathbf{w} | \mathcal{D})
\end{align}
We should emphasize here that the approximate distribution $q(\mathbf{w} | \theta)$ is also related to the data ($\mathcal{D}$). 
Since we want to approximate a probability distribution which is conditionated to $\mathcal{D}$, the approximate distribution shall also depend on the data. 
Consequently, we will (partially) adjust the variational parameters based on the evidence. 
The approximation is realised by minimising the Kullback-Leibler (KL) divergence between $q$ and $P$, what can be seen as an optimisation problem:
\begin{align*}
\theta^{*} &= \arg \min_{\theta} \ KL \left[ q(\mathbf{w} | \theta) || P(\mathbf{w} | \mathcal{D}) \right]
\end{align*}

The KL divergence measures of how the approximate posterior distribution $q$ is different from the theoretical reference posterior distribution $P$.\footnote{The KL divergence is not a true distance metric since it is not symmetric in the two distributions, and does not satisfy the triangle inequality.}
Formally, the KL divergence is defined as follows: 
\begin{align}
KL \left[ q(\mathbf{w} | \theta) || P(\mathbf{w} | \mathcal{D}) \right] = \int q(\mathbf{w} | \theta) \log \frac{q (\mathbf{w} | \theta)}{P(\mathbf{w} | \mathcal{D})}d\mathbf{w} \label{eq:kl_divergence}
\end{align}

Here we face a difficulty because the integral defined by (\ref{eq:kl_divergence}) is intractable. 
Fortunately, we can approximate it as follows. 
We already have approximated the underlying true distribution $P$ with a variational distribution $q$ and we know that we can always sample, i.e. use Monte Carlo methods, from any intractable distribution. 
So, we can approximate our cost function using sampled weights from the variational distribution $q$ while seeing the data. 
This exact step is the essence of Bayes by Backprop: 
First, we approximate the underlying true distribution $P$ with an approximate distribution $q$, the shape of which is parameterized by trainable variational parameters $\theta$, and then sample from that $q$ while seeing data. 
Hence, we arrive at a tractable objective function \citep{blundell2015weight}: 
\begin{align}
\theta^{*} = \arg\min_{\theta} \sum_{i=1}^{n} \log q\left( \mathbf{w}_{(i)} | \theta \right) - \log P\left( \mathbf{w}_{(i)} \right) - \log P\left( \mathcal{D} | \mathbf{w}_{(i)} \right)
\label{eq:elbo}
\end{align}
with sample $\mathbf{w}_{(i)}$ from $q(\mathbf{w}|\theta)$. 
\\

\begin{figure}[H]
\begin{center}
\includegraphics[width = \textwidth]{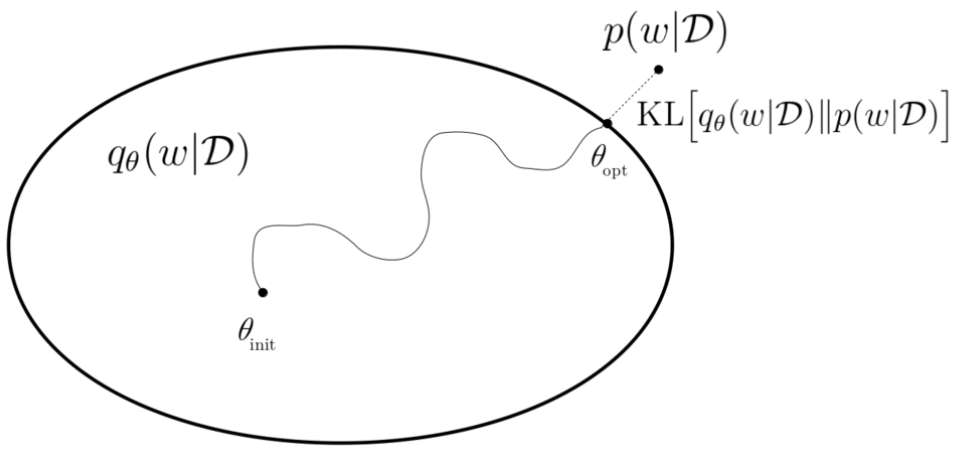}
\caption{Illustration of variational inference. 
We look for an approximate (variational) distribution $q_{\theta}(w | D)$ that is close to the real posterior distribution $P(w | D)$. 
Source: \textcite{laumann2020} } 
\label{fig:variational_inference}
\end{center}
\end{figure}

Figure \ref{fig:variational_inference} tries to give a graphical intuition for this procedure which is given by \textcite{laumann2020}. 

\textbf{Local reparameterization trick}

Let us once again remember that we want to implement the above procedure in a neural network, and therefore must calculate derivatives of the parameters being learnt, i.e. for us, derivatives of distributions. 
For doing so, the local reparameterization trick is deployed which ``moves'' the variational parameters to be learnt, namely the mean $\mu$ and the standard deviation $\sigma$ in case of a Gaussian distribution, out of the distribution function for any weight $\mathbf{w}$ \citep{kingma2015variational}. 
We define $\epsilon$ as a sample from a standard Gaussian distribution, multiply it with the standard deviation $\sigma$ and add the mean $\mu$.
\begin{align*}
\theta &= (\mu, \sigma^{2}) \\
\epsilon &\sim \mathcal{N} \left( 0, 1 \right) \\
\mathbf{w} &= \mu + \sigma \odot \epsilon, 
\end{align*}
where $\odot$ is element-wise (Hadamard) product. 
Doing so, we have these two parameters of interest incorporated in every weight value and can both calculate the derivative of it and re-write it into a probability distribution. 
Our parameters are then updated, i.e. learnt, according to the gradient with respect to the mean and standard deviation parameter. 
According to (\ref{eq:elbo}), let $f(\mathbf{w}, \theta)$ represent the loss for a given weight vector ($\mathbf{w}$) sampled from $q(\mathbf{w} | \theta)$: $f(\mathbf{w}, \theta) = \log q(\mathbf{w} | \theta) - \log P(\mathbf{w}) - \log P(\mathcal{D} | \mathbf{w})$. 
Then, the corresponding gradients are calculated as follows: 
\begin{align}
\Delta_{\mu}  &= \frac{\partial f(\mathbf{w}, \theta)}{\partial \mathbf{w}} + \frac{\partial f(\mathbf{w}, \theta)}{\partial \mu}, \\
\Delta_{\sigma}  &= \frac{\partial f(\mathbf{w}, \theta)}{\partial \mathbf{w}} \epsilon + \frac{\partial f(\mathbf{w}, \theta)}{\partial \sigma}.
\end{align}
With these gradients, we update the variational parameters, $\mu$ and $\sigma$, according to the learning rate ($\alpha$):  
\begin{align*}
\mu \leftarrow \mu - \alpha \Delta_{\mu}, \\
\sigma \leftarrow \sigma - \alpha \Delta_{\sigma}, \\
\theta^{*} = \left( \mu^{*}, \sigma^{*} \right).
\end{align*}

Bayesian models offer a mathematically grounded framework to reason about model uncertainty, but usually come with a prohibitive computational cost. 
While sampling-based variation inference, i.e., Bayes by Backprop really boosted the application of Bayesian neural networks, the additional computation cost is still significant. 
To represent uncertainty with Bayes by Backprop, the number of trainable parameters is doubled for the same network size. 
It is no coincidence that many empirical applications with limited training samples use some kind of ensemble learning instead. 
Different methods of ensemble learning can also generate a distribution of the weights but they do not learn it from the data. 
Intuitively, they provide an approximate solution for the Bayesian inference discussed above.

\subsubsection{Monte Carlo dropout}     \label{sec:mc_dropout}

\textcite{gal2016dropout} introduced Monte Carlo dropout as a new theoretical framework casting dropout training in deep neural networks as approximate Bayesian inference. 
A direct result of this theory gives us tools to model uncertainty with dropout ANNs -- extracting information from existing models that has been thrown away so far. 
This mitigates the problem of representing uncertainty in deep learning without sacrificing either computational complexity or test accuracy. 
As its name suggests, Monte Carlo dropout presumes an underlying ANN that has dropout layers in its architecture. 
So, in the following, we briefly describe these layers and give some intuition about why they have become so popular in recent years. 

Considering its original use case, dropout is a regularization technique for ANNs proposed by \textcite{hinton2012improving}. 
It is realized with a layer in the neural network. 
During training, the dropout layer randomly zeroes some of the elements of the input tensor with probability $p$ before passing it to the next layer. 
The zeroed out neurons do not contribute to the activation of downstream neurons, as they are temporally removed on the forward pass. 
Consequently, any weight updates are not applied to those neuron on the backward pass. 
The zeroed elements are chosen independently for each forward call and are sampled from a Bernoulli distribution which takes the value 1 with probability $1-p$, so that we drop the $i$-th neuron with probability $p$. 
This technique has proven to be an effective technique for regularization and preventing overfitting in ANNs \citep{hinton2012improving}. 

Basically there are two main reasons why switching off some parts of the model might be beneficial. 
First, the information spreads out more evenly across the network. 
If we think about a single neuron somewhere inside our network, then there are presumably a couple of other neurons that provide it with inputs. 
With dropout, each of these input sources can disappear at any time during training. 
Hence, our neuron cannot rely on one or two inputs only, and it has to spread out its weights. 
Intuitively, it is forced to pay attention to all of its inputs. 
As a result, the trained network learns kind of redundancy and becomes less sensitive to input noise. 

The other explanation of dropout's effectiveness is even more important from the point of view of Monte Carlo dropout. 
With dropout, in every training iteration we randomly sample the neurons to be dropped out in each layer (according to that layer’s dropout rate), so a different set of neurons are being dropped out each time. 
Since we train a slightly different ANN architecture in every forward pass, we can think of the final pre-trained model as an averaging ensemble of many different neural networks. 
Each member of this ensemble is trained on one batch of data. 

Until now, we have considered dropout as if we could only use it during training. 
At inference time, when we make predictions with the trained network, we typically do not apply any dropout as we supposedly want to use all the trained neurons and connections. 
However, as it turns out, dropout often improves predictive accuracy at inference (test) time as well. 
Monte Carlo dropout, proposed by \textcite{gal2016dropout} is a clever realization that the use of the regular dropout can be interpreted as a Bayesian approximation of a well-known probabilistic model: the Gaussian process. 
We can treat the many different networks (with different neurons dropped out) as Monte Carlo samples from the space of all available models. 
This provides mathematical grounds to reason about the model's uncertainty and, as it turns out, often improves its performance.

\begin{figure}[H]
\begin{center}
\includegraphics[width = \textwidth]{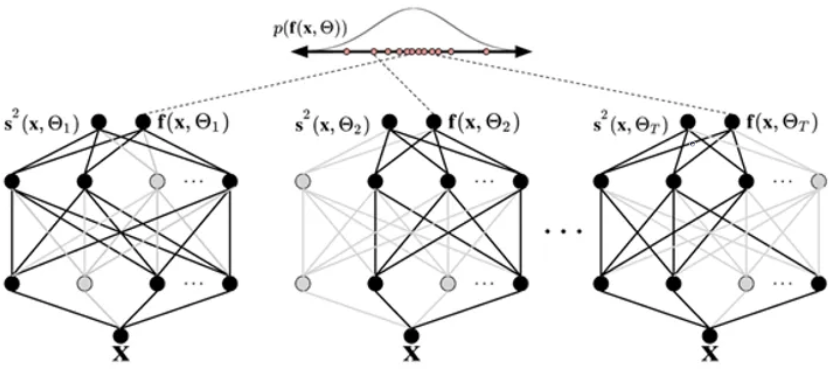}
\caption{Monte Carlo dropout at the inference time (regression): 
Each network shown is acquired by applying dropout to a pre-trained model. 
The same input, $x$, is fed to each ensemble member, and the resulting outputs form a distribution we can use to estimate the uncertainty. 
Source: \textcite{bench2023}. } 
\label{fig:mcdropout_inference}
\end{center}
\end{figure}

To turn on dropout at inference time, we simply need to set training=True to ensure training-like behavior, that is dropping out some neurons. 
This way, each prediction will be slightly different and we may generate as many as we like. 
Each \textit{``version''} of the model at the evaluation stage can be thought of as a sample from the posterior distribution $\mathbb{P}(\theta | D)$, and each resulting output can be thought of as a sample from the predictive distribution $\mathbb{P}(y | x, D)$. 

Similarly to Bayes by Backprop, there are some limitations to using Monte Carlo dropout as well. 
Firstly, it can be computationally expensive due to the need to make multiple predictions for each input from the network. 
This can increase the inference time and may not be practical in real-time applications. 
Secondly, the number of predictions required to obtain a good estimate of the uncertainty can be high, which can further increase the computational cost.

\section{Empirical analysis}    \label{sec:empirical_analysis}

In this Section, we describe the design of the empirical analysis and then report the results. 
First, we distinguish three different intra-quarterly information sets or nowcasting scenarios based on which nowcasts are conducted and evaluated. 
Then, we define how density and point nowcasts are generated for each model (algorithm) and present the estimation (training) procedure.
After that, we discuss the results for the benchmark DFM and the two DL algorithms.

\subsection{Information sets and density nowcasts}

The full dataset ranges from 1960:Q1 to 2022:Q4, and it contains regressor (feature) observations for every month and target observations for every third month (i.e., for every quarter). 
We assume that quarterly GDP measurements are available for months M3, M6, M9, and M12 within a year.\footnote{FRED data releases associate these quarterly measurements with timestamps M1, M4, M7, and M10. However, using these timestamps would be counter-intuitive in an empirical analysis centered around nowcasting. 
Following the literature, we assume that quarterly GDP measurements are published at the end of the current quarter. In reality, the first measurements are usually available only a few months after the end of the current quarter.}  
Our evaluation period ranges from 2012:Q1 to 2022:Q4, containing 132 measurements for the monthly regressors and 44 observations for the target variable. 
Consequently, we will have 44 test sequences for evaluation. 
Regarding the characteristics of the target variable, the evaluation set includes periods of both balanced growth (2012 -- 2019) and high economic turbulence (2020 -- 2021). 
We use real-time vintages of the monthly FRED-MD database according to the test period. 
Accordingly, the first FRED-MD vintage used in the empirical analysis was released in 2012:M1. 
Depending on the information set based on which nowcasts are generated, we use every third FRED-MD vintage until the end of the evaluation period.\footnote{The real-time FRED-MD vintages are openly available and can be downloaded from \url{https://research.stlouisfed.org/econ/mccracken/fred-databases/}}

During the empirical analysis, we distinguish three different information sets or \textit{nowcasting scenarios} based on which nowcasts are conducted and evaluated.  
Since we use the real-time monthly vintages of the FRED-MD database, these information sets (nowcasting scenarios) are related to the publication schedule of those vintages. 
Concretely, the first \textit{'1-month'} information set ($\mathbf{\Omega}_{m_{1}}$) is related to FRED-MD vintages that are released in the first month of a given quarter. 
Within a year, those vintages are associated with timestamps M1, M4, M7 and M10. 
In line with the above, we let a quarter $q$ be dated by its last month so that it is associated with one of the timestamps M3, M6, M9, and M12. 
Consequently, the relation between quarterly and monthly time indices can be expressed as $t = 3q$.\footnote{For example, the first quarter ($q=1$) refers to the third month ($t=3$) of the sample, and so on.} 
Hence, the GDP growth in quarter $q$ is denoted by $y_{3q}$ using monthly time indices. 
Then, $\mathbf{\Omega}_{m_{1}}$ consists of input sequences where target observations are associated with such regressor vectors $(\mathbf{X}_{m_{1}})$ that contain monthly indicator data until the first intra-quarterly month: $m_1 = 3(q-1)+1$.
Formally, $\mathbf{\Omega}_{m_{1}} = \bigcup\displaylimits_{q} \{ \left( \mathbf{X}_{m_{1}}, y_{3q} \right) \}$, where $\left( \mathbf{X}_{m_{1}}, y_{3q} \right)$ is an input sequence composed by a quarterly target measurement ($y_{3q}$) and its associated regressor vector $\mathbf{X}_{m_{1}}$, both indexed on a monthly frequency. 
The regressor vector $\mathbf{X}_{m_{1}}$ is defined as $\mathbf{X}_{m_{1}} = \left[ \mathbf{x}_{m_{1}-l+1}, \dots, \mathbf{x}_{m_{1}} \right]$, where $l$ stands for the length (measured in months) of the input sequences. 

Similarly, the second \textit{'2-month'} information set or nowcasting scenario ($\mathbf{\Omega}_{m_{2}}$) is related to those FRED-MD vintages released during the second intra-quarterly month. 
Following the above definitions, $\mathbf{\Omega}_{m_{2}} = \bigcup\displaylimits_{q} \{ \left( \mathbf{X}_{m_{2}}, y_{3q} \right) \}$, where $m_2 = 3(q-1) + 2$ is the second intra-quarterly month. 
Finally, the third \textit{'3-month'} information set or nowcasting scenario ($\mathbf{\Omega}_{m_{3}}$) corresponds to those FRED-MD vintages released in the last month of a given quarter: 
$\mathbf{\Omega}_{m_{3}} = \bigcup\displaylimits_{q} \{ \left( \mathbf{X}_{m_{3}}, y_{3q} \right) \}$, where $m_3 = 3(q-1) + 3$ stands for the last month of the current quarter. 
Together, the distinguished information sets form a three-step nowcasting window along which density nowcasts are generated by the different models and algorithms. 
In machine learning, the separate set of training sequences (examples) is also commonly defined. 
Following the convention, we define it as: $\mathcal{D}_{m} = \{ (\mathbf{X}_{m}; y_{3q}) \}_{q} \subset \mathbf{\Omega}_{m}$, $m \in \{ m_1, m_2, m_3 \}$. 

It is worth noting that the definitions of those information sets reflect a somewhat optimistic, ideal situation of monthly data releases. 
As is common in nowcasting, we should deal with missing regressor data at the end of the estimation window, i.e., the ragged edges. 
For example, the monthly FRED-MD vintages corresponding to the first intra-quarterly month typically have missing values for that specific month, i.e., for $m_1$. 
Usually, they contain valid regressor observations only until the preceding month. 
We can also see this pattern for those FRED-MD vintages corresponding to the 2-month and the 3-month information sets. 
While the DFM  handles those missing values straightforwardly via the Kalman filter, applying the two DL algorithms requires some \textit{adequate} extrapolation routine. 
We extrapolate for the ragged edges by applying univariate AR(1) forecasts of each single monthly regressor. 
We also investigated a more sophisticated extrapolation technique based on univariate ARMA(p,q) forecasts, but this resulted in slightly less accurate nowcasts for both Bayes by Backprop and Monte Carlo dropout. 

Based on the above, our density nowcasting problem involves a single quarterly outcome variable (indexed on a monthly frequency as $y_{3q}$), and a monthly regressor vector ($\mathbf{X}_{m}$) that includes conditioning variables available up through the $m$-th intra-quarterly month, $m \in \{ m_1, m_2, m_3 \}$. 
Then, we can define the density nowcast for $y_{3q}$, as the conditional (predictive) distribution of $y_{3q}$, given regressor vector $\mathbf{X}_m$ and pre-trained model $\mathcal{M}$ \citep{elliott2016economic}: 
\begin{align}   \label{eq:density_nowcast}
\mathcal{P}(y_{3q}) = P( y_{3q} | \mathbf{X}_m, \mathcal{M}), 
\end{align}
where $\mathcal{M}$ is trained with those input sequences included in $\mathcal{D}_m$ according to the underlying information set. 
To evaluate the accuracy of the predictive densities, we also define the point nowcast for $y_{3q}$ with its conditional expected value: 
\begin{align}   \label{eq:point_nowcast}
\hat{y}_{3q} = E ( y_{3q} | \mathbf{X}_m, \mathcal{M}). 
\end{align}

In the case of the DFM, the conditional distribution of the measurement (target) variable is assumed to be normal, and it is derived analytically in a two-step nowcasting procedure. 
Based on equations (\ref{eq:dfm_mean}) and (\ref{eq:dfm_variance}) the predictive density of $y_q$ and its conditional expected value is generated as follows:\footnote{It is important no note that notations used for DFM are slightly different from that used for the two DL algorithms. 
As common in the context of state space models, the information set is denoted by $Y$ \citep{harvey1990forecasting}.} 
\begin{align}
P ( y_{3q} | Y_{m} ) &= \mathcal{N}\left( E\left[ y_{3q} | Y_{m} \right], Var\left[ y_{3q} | Y_m \right] \right) 
= \mathcal{N}\left( \hat{\mu}_{3q | m}, \ \hat{\sigma}^{2}_{3q | m} \right), \label{eq:dfm_density} \\
E ( y_{3q} | Y_m ) &= \hat{\mu}_{3q | m} 
\label{eq:dfm_point}.
\end{align}

By contrast to the DFM, the two DL algorithms described above cannot generate predictive densities in an analytic (closed) form. 
This follows from the fact that these algorithms do not impose any explicit parametric restriction (assumption) on the conditional distribution of $y_q$. 
Instead, in the case of Bayes by Backprop, we place a prior upon the weights of the underlying ANN. 
Since we assume a multivariate Gaussian with diagonal covariance matrix as the prior of the weights, and Bayes by Backprop can only change modify (optimize) the parameters of the prior distribution, the (approximate) posterior of the weights will also be a multivariate Gaussian with mean vector $\mu^{*}$ and diagonal covariance matrix $\Sigma^{*}$: $q(\mathbf{w} | \theta^{*}) = \mathcal{N}(\mu^{*}, \Sigma^{*})$. 
Taking $n$ random sample from this distribution results in $n$ different weight vectors as follows: 
\begin{align*}
\mathbf{w}_{(1)}, \mathbf{w}_{(2)}, \mathbf{w}_{(3)}, \dots, \mathbf{w}_{(n)} \sim \mathcal{N}( \mu^{*}, \Sigma^{*})
\end{align*}

According to equation (\ref{eq:density_nowcast}), $P( y_{3q} | \mathbf{X}_m, \mathcal{D}_m)$ stands for the conditional (predictive) distribution of the target variable. 
As we cannot take an infinite amount of samples from the posterior, we approximate the probability density function with a kernel density estimator ($\Psi$) as follows:\footnote{For simplicity, we drop the time indices for $y_{3q}$ $\mathbf{X}_m$ $\mathcal{D}_m$ in the following equations, making them easier to read. 
Similarly to the DFM, density and point nowcasts generated by the two DL algorithms can be defined at each step of the nowcasting window.}
\begin{align}
P(y | \mathbf{X}, \mathcal{D}) &= \int\displaylimits_{\mathbf{w}} P\left( y | \mathbf{X}, \mathbf{w} \right) \cdot P\left( \mathbf{w} | \mathcal{D} \right) \ d\mathbf{w} 
\approx \Psi \left( \{( y (\mathbf{X}, \mathbf{w}_{(i)}) \}_{\mathbf{w}_{(i)} 
\sim q\left( \mathbf{w} | \theta^{*} \right))} \right), \label{eq:bbb_density} \\ 
E(y | \mathbf{X}, \mathcal{D}) &\approx \frac{1}{n} \sum\displaylimits_{i=1}^{n} \  y( \mathbf{X}, \mathbf{w}_{(i)} ), \label{eq:bbb_point}
\end{align}
where $\Psi()$ stands for the kernel density estimator of the posterior probability density function \citep{parzen1962estimation, rosenblatt1956kde}. 

Generating density nowcasts with Monte Carlo dropout also relies on sampling. 
While in Bayes by Backprop sampling meant to draw weights from the variational (approximate) posterior distribution, here we sample multiple ($n$) dropout masks drawn from a multivariate Bernoulli distribution $\mathbb{B}(1-p)$. 
In line with the earlier, $p$ denotes the dropout rate. 
The sampled dropout mask $\mathbf{z}_{(i)}$ uniquely identifies the $i$-th member of the ensemble, $\mathbf{z_{(i)}} \in \{0,1\}^{M}$, where $M$ is the number of the neurons. 
In Monte Carlo dropout, members will have the same number of weights ($N$) as the ensemble but randomly filled with zeros, proportionally to the dropout rate. 
Those weights not affected (selected) by the dropout mask are shared between the member and the ensemble. 
If $\mathbf{w^{*}}$ denotes the optimal weight vector learned by the ensemble, and $\mathbf{w}_{(i)}^{*}$ stands for the weights corresponding to the ($i$)-th ensemble member, then the following equations describe how these weight vectors are related: 
\begin{align*}
\mathbf{w^{*}} &= w_{1}^{*}, w_{2}^{*}, w_{3}^{*}, \dots, w_{N-2}^{*}, w_{N-1}^{*}, w_{N}^{*},  
\\
\mathbf{w_{(i)}^{*}} &= \mathbf{w^{*}} \odot \gamma(\mathbf{z}_{(i)}) \qquad \mathbf{z}_{(i)} \sim \mathbb{B}( 1-p ), 
\end{align*}
where $\gamma(\mathbf{z})_{j}$ is 0 if and only if the $j$-th weight of the network is in the output of the $i$-th neuron and $z_{i}=0$, while it is 1 if and only if $z_{i}=1$. 
After sampling $n$ different dropout masks, i.e., selecting $n$ different members from the ensemble, we can estimate the predictive distribution of GDP growth as follows:  
\begin{align}
P(y | \mathbf{X}, \mathcal{D}) &= \int\displaylimits_{\mathbf{z} \in \{0,1\}^{M}} y\left( \mathbf{X}, \mathbf{w}^{*} \odot \gamma(\mathbf{z}) \right) 
\cdot \mathbb{B}\left( \mathbf{z} \right) d\mathbf{z} 
\approx \Psi \left( \{( y (\mathbf{w}^{*} \odot \gamma(\mathbf{z}_{(i)}), \mathbf{X}) \}_{\mathbf{z}_{(i)} 
\sim \mathbb{B}( 1-p )} \right), 
\label{eq:mcdropout_density} \\
E(y | \mathbf{X}) &\approx \frac{1}{n} \sum\displaylimits_{i=1}^{n} \  y( \mathbf{X}, \mathbf{w}_{(i)}^{*} ). 
\label{eq:mcdropout_point}
\end{align}

The underlying ANN used with those DL algorithms is estimated (trained) based on a rolling estimation (training) window, where the initial estimation (training) window ranges from 1960:Q1 to 2009:Q4. 
So, the size of the rolling window contains 200 training sequences: 
600 measurements for the monthly regressors (indicators) and 200 observations for quarterly GDP growth, associated with timestamps M3, M6, M9, M12. 
The corresponding initial validation window (set) used for early stopping is from 2010:Q1 to 2011:Q4. 
Following the recommendation of the literature, we keep a small fraction of the data aside from training and use it for early stopping: 
8 observations for the target variable and 24 for the explanatory variables (\citep{amari1997asymptotic}). 
Accordingly, first density nowcasts are generated for 2012:Q1 by the different models. 
The number of increments between successive estimation windows is 1 quarter (3 months). 
Based on equations (\ref{eq:dfm_density}), (\ref{eq:bbb_density}), and (\ref{eq:mcdropout_density}), we generate the series of density nowcasts for the evaluation period where $q \in \{ $2012:Q1$, \dots $2022:Q4$ \}$. 

As for the Bayesian neural network, we use the evidence lower bound (ELBO) defined by equation (\ref{eq:elbo}) as the loss (criterion) function, where the reconstruction error is measured by the mean squared error of the point nowcasts. 
Training the network with Monte Carlo dropout only relies on the reconstruction loss (i.e., evidence), so we use the mean squared error of the point nowcasts.  
The predictive accuracy of the two DL algorithms is evaluated relative to a naive constant growth model for GDP and our benchmark DFM specification (see: Section \ref{sec:dfm}). 
Uncertainty of nowcasts is measured by the standard deviation of the (empirical) predictive distributions.

\subsection{Results for the benchmark DFM}

In this section, we present the results for the two-step nowcasting procedure based on our benchmark DFM specification defined by (\ref{eq:dfm_measurement})--(\ref{eq:dfm_factor}). 
Following section \ref{sec:dfm}, the two-step nowcasting procedure refers to estimating two separate state space models. 
First, the common factor is estimated based on the DFM, which is then used as an exogenous regressor for the unobserved monthly growth in a second state space model. 
This latter state space model is specified by state equation (\ref{eq:ssm_state}) and measurement equation (\ref{eq:ssm_measurement}), and it generates nowcasts for quarterly GDP growth. 
Both of these models' parameters are estimated with an expectation-maximization algorithm, iteratively maximizing the log-likelihood given by \ref{eq:ssm_loglikelihood}. 
Table \ref{tab:sspace_params} shows the outcome of the maximum likelihood estimation for the second state space model directly used for nowcasting. 

\begin{table}[H]
\centering
\caption{Maximum likelihood estimation of the state space model (\ref{eq:ssm_state})-(\ref{eq:ssm_measurement}) along the rolling estimation window.}
\centering
\def\arraystretch{1.2}
\begin{tabular}{l | r r r l}
& Coefficient & Std. Error & t Statistics & p-Value \\
\hline\hline
$c$ & $0.257 \rightarrow 0.228$ & $0.014$ & $16.654$ & $0^{***}$ \\ 

$\varphi$ & $0.201 \rightarrow 0.372$ & $0.019$ & $21.314$ & $0^{***}$ \\ 

$\sigma^{2}_{\varepsilon}$ & $0.198 \rightarrow 0.211$ & $0.015$ & $24.831$ & $0^{***}$ \\
\bottomrule
\end{tabular}
\caption*{\begin{footnotesize}
\textbf{Notes:} The stars denote statistical significance at the 10\%($^{*}$), 5\%($^{**}$) and 1\%($^{***}$) level respectively. 
\end{footnotesize}}
\label{tab:sspace_params}
\end{table}

Table \ref{tab:sspace_params} shows that the estimated coefficients are significant along the whole rolling window. 
The comovement of quarterly growth and the common factor, captured by parameter $\varphi$, by the end of the rolling window. 
That is in line with the intuition because comovement (correlation) across economic indicators and GDP growth is usually found to be stronger during and following recessions \citep{labonne2020capturing}. 

\begin{figure}[H]
\begin{center}
\includegraphics[width = \textwidth]{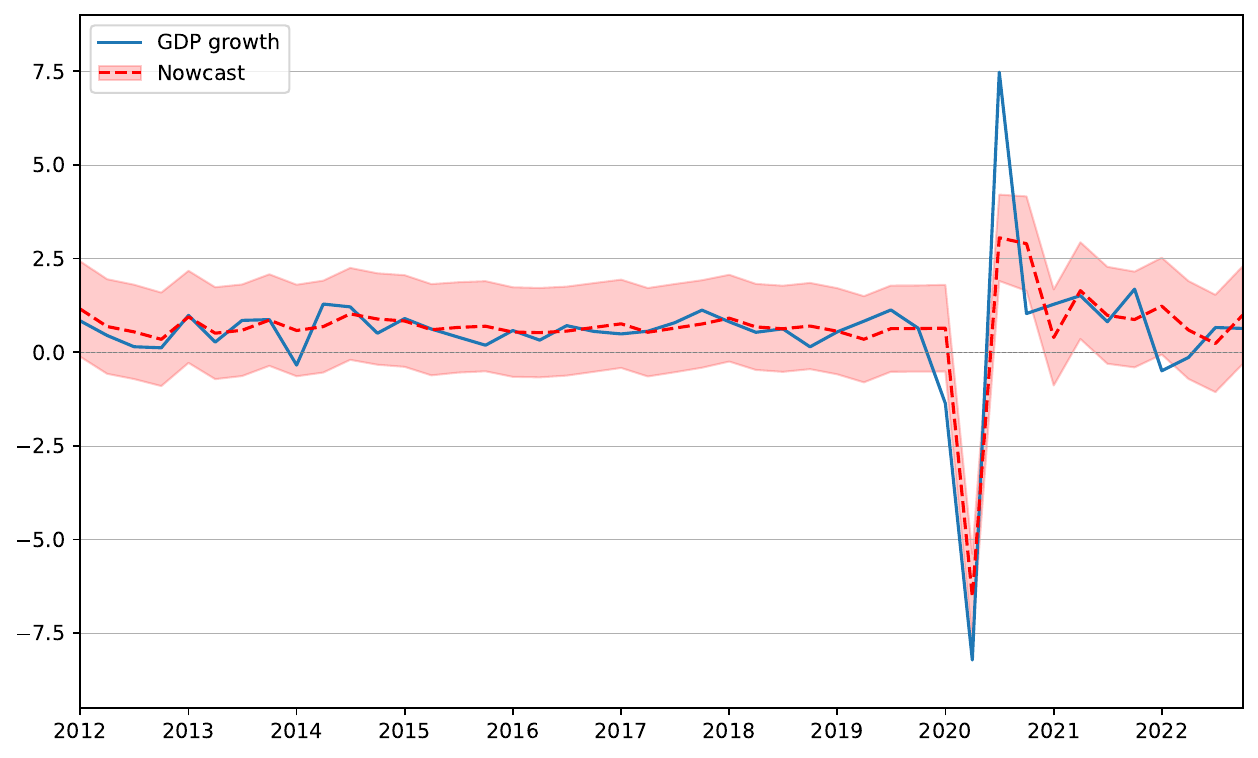}
\caption{Interval nowcasts generated by the DFM. 
Mean of the predictive distribution surrounded by a two standard deviation uncertainty range (95\% prediction interval). 
Evaluation period: 2012:Q1 -- 2022:Q4. 
Source: Own calculations based on FRED-MD.}
\label{fig:dfm_interval}
\end{center}
\end{figure}

Figure \ref{fig:dfm_interval} below plots the series of interval nowcasts generated by the DFM during the evaluation period against the actual GDP growth. 
The mean (median) of the predictive distribution is surrounded by a two standard deviation uncertainty range. 
Since the predictive (conditional) distribution of GDP growth is assumed to be Gaussian, we get a 95\% prediction interval for the DFM. 

As Figure \ref{fig:dfm_interval} shows, the benchmark DFM fits the data generally well during the evaluation period. 
That is especially true for the periods of balanced economic growth, i.e., between 2012 and 2019. 
Basically, that is what we can expect from a linear model. 
Furthermore, the DFM also generates fairly accurate nowcasts during the sharp V-shaped recession caused by the COVID-19 pandemic. 
Similarly to the results of \textcite{loermann2019nowcasting}, we see that model captures the downturn in 2020:Q2 and it only really misses the quick rebound in the successive quarter. 
Accompanied by 2022:Q1, these are the only quarters where the actuals fall outside the two standard deviation prediction interval. 
Based on Figure \ref{fig:dfm_interval}, we should also mention that nowcasting accuracy of the DFM gradually weakens as we approach the end of the test period. 
That seems intuitive, because economic growth has become persistently more volatile after the COVID recession. 
Nonetheless, the DFM generates an RMSE of $0.937$ and an MAE of $0.519$ at the end of the nowcasting window. 
To put these values into perspective, Table \ref{tab:dfm_accuracy} shows a relative RMSE of $0.331$ and a relative MAE of $0.467$ compared to a naive constant growth model for GDP. 
As Table \ref{tab:dfm_accuracy} shows, the DFM consistently and significantly beats the naive benchmark model in terms of nowcasting accuracy. 
Regarding RMSE evaluation, it significantly outperforms the naive benchmark model based on the second (2-month) and third (3-month) information sets. 
Regarding MAE evaluation, the DFM beats the naive benchmark model at a 5\% significance level at every step of the nowcasting window. 

\begin{table}[H]
\centering
\caption{Accuracy of point nowcasts generated by the DFM. 
RMSE and MAE evaluation. 
The evaluation period ranges from 2012:Q1 to 2022:Q4.}
\def\arraystretch{1.0}
\begin{tabular}{l | l l | l l | l l}
{} & \multicolumn{2}{c|}{$\Omega_{m_{1}}$} & \multicolumn{2}{c|}{$\Omega_{m_{2}}$} & \multicolumn{2}{c}{$\Omega_{m_{3}}$} \\
& RMSE & MAE & RMSE & MAE & RMSE & MAE \\
\hline\hline
DFM (Statsmodels) &  $1.420$ &  $0.670$ &  $0.918$ &  $0.490$ &  $0.936$ &  $0.523$ \\
{} &    $0.501$ &   $0.604^{**}$ &  $0.324^{*}$ &   $0.441^{**}$ &  $0.330^{*}$ &    $0.471^{**}$ \\
\bottomrule
\end{tabular}
\caption*{\begin{footnotesize}
\textbf{Notes:} This table reports the results of RMSE and MAE evaluation for the point nowcasts (conditional mean) generated by the DFM. 
The first line shows absolute RMSE and MAE values. 
The second line reports relative RMSE and MAE values relative to a naive constant growth model for GDP. 
A value below one indicates that the competitor model beats the naive benchmark model. 
The stars denote statistical significance at the 10\%($^{*}$), 5\%($^{**}$) and 1\%($^{***}$) level of the one-sided \textcite{diebold1995comparing} test. 
Columns are related to the different information sets in the nowcasting window: e.g., $\Omega_{m_{1}}$ refers to the first (1-month) information set. 
\end{footnotesize}}
\label{tab:dfm_accuracy}
\end{table}

\begin{figure}[H]
\begin{center}
\includegraphics[width = \textwidth]{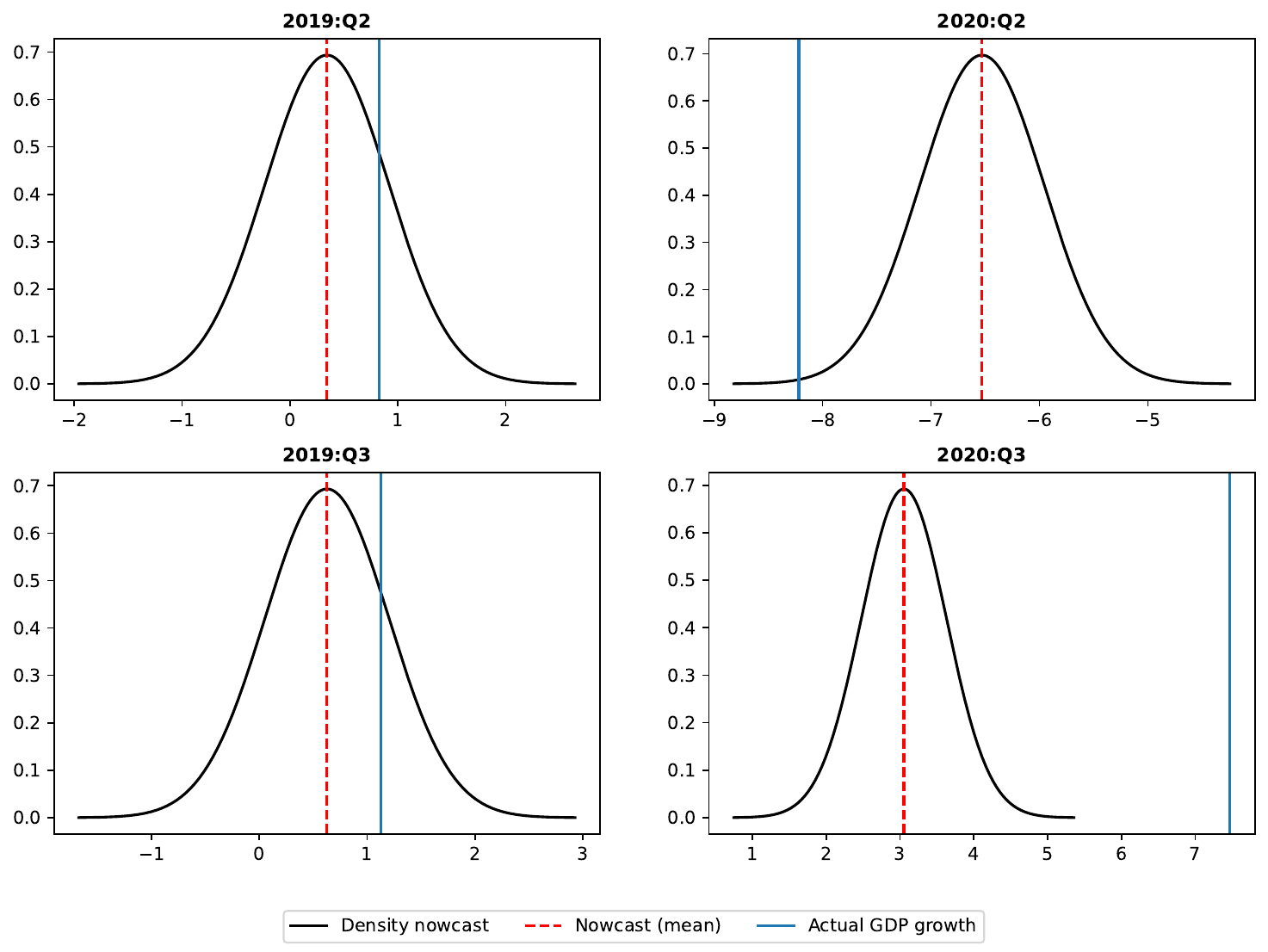}
\caption{Density nowcasts generated by the DFM. 
Nowcasts are conducted based on the 3-month information set ($\Omega_{m_{3}}$), i.e., at the end of the nowcasting window. 
Source: Own calculations based on FRED-MD.} 
\label{fig:dfm_density}
\end{center}
\end{figure}

As it follows from the specification of the underlying state space models, the DFM generates normally distributed density nowcasts for $y_{3q}$ (see: equation \ref{eq:dfm_density}). 
Figure \ref{fig:dfm_interval} shows that assumptions incorporated in the DFM's specification seem legit for the majority of the evaluation period. 
and the uncertainty range also works as intended: 
Apart from a few \textit{problematic} quarters, actual growth falls inside the 95\% prediction interval. 
Unfortunately, though, the DFM misses those critical quarters where timely detection of a large shock or structural break would be particularly important. 
While its parsimonious, linear specification lends a very good generalization capability to the DFM for the majority of periods, it is restricted from learning a time-varying variance or shape (skew) parameter for the predictive density. 
This problem is highlighted in Figure \ref{fig:dfm_density} below where we see density nowcasts generated by the DFM for four different quarters. 

Figure \ref{fig:dfm_density} below shows density nowcasts generated by the DFM for four different quarters: 
2019:Q2 (upper-left panel), 2019:Q3 (lower-left), 2020:Q2 (upper-right) and 2020:Q3 (lower-right). 

Figure \ref{fig:dfm_density} shows that, the benchmark DFM can dynamically adjust the location parameter ($\mu$) of the predictive distribution. 
This can be clearly seen during the V-shaped COVID recession: 
Here the DFM adjusts the conditional mean from $-6.843$ to $3.072$ between consecutive quarters. 
The actual GDP growth was $-8.22$ and $7.47$ percent in 2020:Q2 and in 2020:Q3 respectively. 
Along with that, Figure \ref{fig:dfm_density} also shows that the benchmark DFM cannot dynamically update the scale (standard deviation, $\sigma$) and the shape (skew) of the predictive nowcast distribution. 
Considering these limitation, the first one seems particularly important. 
Predicting constant scale basically means that the model cannot differentiate the periods of low uncertainty (e.g., 2019:Q3) from those periods of high uncertainty (e.g., 2020:Q3). 
Thus the model does not provide reliable information about the expected nowcast error or any suggestions about possible regime changes. 

Although our paper considers the DFM mainly as a benchmark, we briefly mention the possible extensions of the linear specification. 
First, it is possible to introduce time-variation in the variance of the common factors, which would lead to a state space model with stochastic volatility \citep{antolin2017tracking, marcellino2016short}. 
While the extension of the specification seems straightforward, the Kalman filter cannot be used any longer as an unbiased estimator. 
Similarly, the normality of the innovations can also be generalized in non-Gaussian state space models. 
These extensions, however, come at high computational costs, and non-Gaussian state space models often result in an inferior performance in terms of accuracy of the point nowcasts. 
It is no coincidence that score driven models have received much more attention in the literature instead of non-Gaussian state space models \citep{creal2014observation, delle2017adaptive, gorgi2019forecasting}. 

Summarizing the above, the benchmark DFM dynamically and effectively adjusts the conditional mean of the predictive distribution. 
By doing so, it fits the data generally well during the evaluation period, especially in periods of balanced economic growth. 
At the same type, it fails to change the the scale and the shape of the predictive distribution, providing no timely signals about potential regime switches or structural breaks. 
In the followings we evaluate the density nowcasts generated by the two deep learning algorithms in question: Bayes by Backprop and Monte Carlo dropout.

\subsection{Results for the DL algorithms}

This section presents the results for Bayes by Backprop and Monte Carlo dropout. 
First, Table \ref{tab:anns_accuracy} reports the relative RMSE and MAE values of the point nowcasts generated by these \textit{competitor} algorithms. 

\begin{table}[H]
\centering
\caption{Accuracy of point nowcasts generated by the competitor DL algorithms. RMSE and MAE evaluation relative to a naive constant growth model for GDP and the benchmark DFM. 
The evaluation period ranges from 2012:Q1 to 2022:Q4.} 
\centering
\def\arraystretch{1.0}
\begin{tabular}{l | l l | l l | l l}
{} & \multicolumn{2}{c|}{$\Omega_{m_{1}}$} & \multicolumn{2}{c|}{$\Omega_{m_{2}}$} & \multicolumn{2}{c}{$\Omega_{m_{3}}$} \\
{} & RMSE & MAE & RMSE & MAE & RMSE & MAE \\
\toprule
Bayes by Backprop (mean)   &  $0.661$ &  $0.702^{*}$ &  $0.297^{*}$ &  $0.470^{**}$ &  $0.202^{*}$ &  $0.367^{**}$ \\
{}   &  $1.321$ &  $1.163$ &  $0.919$ &   $1.066$ &  $0.613$ &   $0.778$ \\
\hline
Bayes by Backprop (median) &  $0.660$ &  $0.699^{*}$ &  $0.299^{*}$ &  $0.474^{**}$ &  $0.202^{*}$ &  $0.370^{**}$ \\
{} &  $1.319$ &  $1.159$ &  $0.925$ &   $1.075$ &  $0.612$ &   $0.784$ \\
\hline
\hline

MC dropout (mean) &  $0.448$ &  $0.507^{*}$ &  $0.219^{*}$ &  $0.377^{**}$ &  $0.196^{*}$ &  $0.365^{**}$ \\
{}  &  $0.895$ &  $0.839$ &  $0.678$ &   $0.855$ &  $0.593^{*}$ &   $0.776$ \\
\hline
MC dropout (median)     &  $0.449$ &  $0.496^{*}$ &  $0.206^{*}$ &  $0.361^{**}$ &  $0.191^{*}$ &  $0.353^{**}$ \\
{}  &  $0.897$ &  $0.821$ &  $0.638$ &   $0.819$ &  $0.577^{*}$ &   $0.750$ \\
\bottomrule
\end{tabular}
\caption*{\begin{footnotesize}
\textbf{Notes:} This table reports the results of RMSE and MAE evaluation for the point nowcasts generated by Bayes by Backprop and Monte Carlo dropout. 
Point nowcasts are defined in two ways: as the mean and the median of the empirical predictive distribution. 
Relative RMSE and MAE values are calculated compared to a naive constant growth model for GDP and the benchmark DFM. 
A value below one indicates that the competitor model beats the naive benchmark model. 
The stars denote statistical significance at the
10\%($^{*}$), 5\%($^{**}$), and 1\%($^{***}$) level of the one-sided \textcite{diebold1995comparing} test. 
Columns are related to the different information sets in the nowcasting window: e.g., $\Omega_{m_{1}}$ refers to the first (1-month) information set. 
\end{footnotesize}}
\label{tab:anns_accuracy}
\end{table}

First, Table \ref{tab:anns_accuracy} indicates that nowcasting accuracy improves towards the end of the nowcasting window. 
Similarly to DFM, the improvement in accuracy is particularly noticeable between the 1-month and 2-month information sets. 
Given that most monthly indicators have missing observations for the first intra-quarterly month in $X_{m_{1}}$, 
At the beginning of the nowcasting window, models are usually restricted to extrapolate based on the previous quarters' economic state. 
So they cannot incorporate the information content of higher frequency indicators that might be relevant to the current quarter. 
Accuracy is basically the same regardless of the aggregation method used for obtaining point nowcasts. 
Thus, our definition for $y_q$ seems appropriate and robust considering this predictive analysis. 

Regarding nowcasting accuracy, Bayes by Backprop significantly outperforms the naive benchmark model at each step of the nowcasting window. 
In terms of RMSE evaluation, it beats the naive benchmark model on a 10\% significance level in the 1-month and 2-month nowcasting scenarios. 
Regarding MAE evaluation, the performance advantage is even more pronounced: 
It is already significant on a 10\% level in the 1-month nowcasting scenario and on a 5\% level in the 2-month and 3-month nowcasting scenarios. 
Compared to the DFM, we see a much more varied picture of nowcasting performance. 
At the beginning of the nowcasting window, accuracy of Bayes by Backprop is slightly worse than that of the DFM. 
Nowcasting accuracy balances out in the 2-month nowcasting scenario, which is indicated by relative RMSE and MAE values close to one. 
Bayes by Backprop comes on top in the last step of the nowcasting window, achieving a relative RMSE of $0.613$ and a relative MAE of $0.778$. 
While the values of those relative evaluation metrics are clearly below one, the performance advantage of Bayes by Backprop is largely due to the better prediction of the COVID-19 recession. 
Otherwise, nowcasting accuracy is very similar for the rest of the evaluation period. 

Monte Carlo dropout also consistently beats the naive benchmark model at each step of the nowcasting window. 
The significance of the performance advantage is very similar to what we have seen for Bayes by Backprop. 
However, it is important to note that relative RMSE and MAE values are generally smaller at each step of the nowcasting window. 
Monte Carlo dropout maintains this advantage also in comparison with the DFM. 
Based on Table \ref{tab:anns_accuracy}, nowcast generated by Monte Carlo dropout are more accurate than those produced by the DFM in every nowcasting scenario. 
Furthermore, the performance advantage is significant on a 10\% level in the last step of the nowcasting window. 

So far, we have only assessed the accuracy of the nowcasts generated by the two DL algorithms. 
In the following, we will focus on other characteristics of the density nowcasts. 
Beyond accuracy, there are much fewer standard approaches to measure the \textit{``goodness''} of a predictive density. 
Notwithstanding the fact that predictive distributions obtained from Bayes by Backprop and Monte Carlo dropout are non-parametric, it is difficult to choose an adequate reference distribution in the first place. 
Consequently, following the literature, we will rely on the graphical and descriptive evaluation of the predictive densities. 
As presented by \textcite{labonne2020capturing}, sudden changes in the scale (variance) of the density reflect an increased (decreased) uncertainty of nowcasts. 
From this point of view, we are interested in seeing whether the DL algorithms can learn the behavior of a stochastic volatility model? 
Furthermore, the shape of the predictive densities might also provide valuable additional information for decision-makers. 
For example, a predictive density that is significantly skewed towards negative (positive) values indicates an increased probability of negative (positive) nowcast errors.  
Obviously, periods of the V-shaped COVID recession deserve special attention from this point of view. 
So, we will analyze density nowcasts for those periods separately at every step of the nowcasting window. 
However, before that, let us present the results for the whole evaluation period.

\subsubsection{Results for Bayes by Backprop}

Figure \ref{fig:bbb_interval} below shows the interval nowcasts generated by the 1D CNN using Bayes by Backprop. 
As discussed earlier, ANNs cannot derive predictive distributions in analytic form, even when using Bayes by Backprop or Monte Carlo dropout. 
Consequently, Figure \ref{fig:bbb_interval} also plots a series of boxplots illustrating how the empirical predictive distribution has evolved during the evaluation period. 

\begin{figure}[H]
\begin{center}
\includegraphics[width = \textwidth]{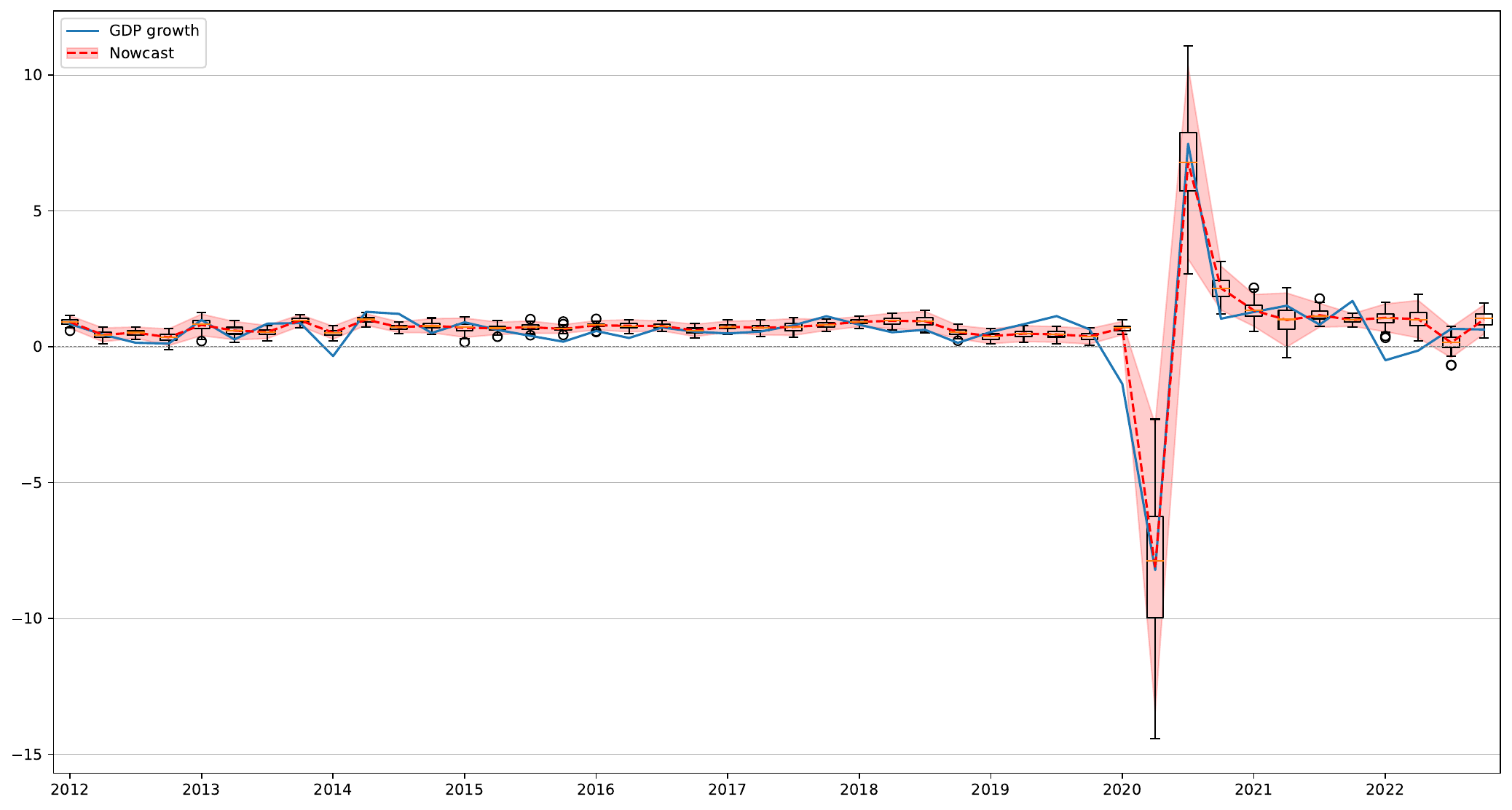}
\caption{Interval nowcasts generated by Bayes by Backprop. 
To create prediction intervals, we measured two times the standard deviation of the predictive distribution around the mean (red dashed line). 
Source: Own calculations based on FRED-MD.}
\label{fig:bbb_interval}
\end{center}
\end{figure}

Figure \ref{fig:bbb_interval} shows that interval nowcasts generated by Bayes by Backprop are quite different from what we have seen for the DFM in Figure \ref{fig:dfm_interval}. 
As Figure \ref{fig:mcdropout_interval} indicates, Bayes by Backprop replicates the behavior of a stochastic volatility model, which results in a much narrower prediction intervals during the pre-COVID part of the evaluation period. 
In this period, the algorithm performs on par with the DFM in terms of accuracy, but actual GDP growth falls outside the two standard deviation prediction interval in several quarters. 
In return, it handles the extreme values of the COVID recession much better and more adaptively. 
It is also apparent that the uncertainty range becomes wider during and right after the COVID crisis. 
Prediction intervals shrink back to their original size by 2022, indicating that economic activity has reverted to its pre-pandemic state. 
Figure \ref{fig:bbb_density} below shows the empirical nowcast distribution (i.e., histograms) generated by the Bayesian neural network for four different quarters: 
2019:Q2 (upper-left panel), 2019:Q3 (lower-left), 2020:Q2 (upper-right) and 2020:Q3 (lower-right). 

\begin{figure}[H]
\begin{center}
\includegraphics[width = \textwidth]{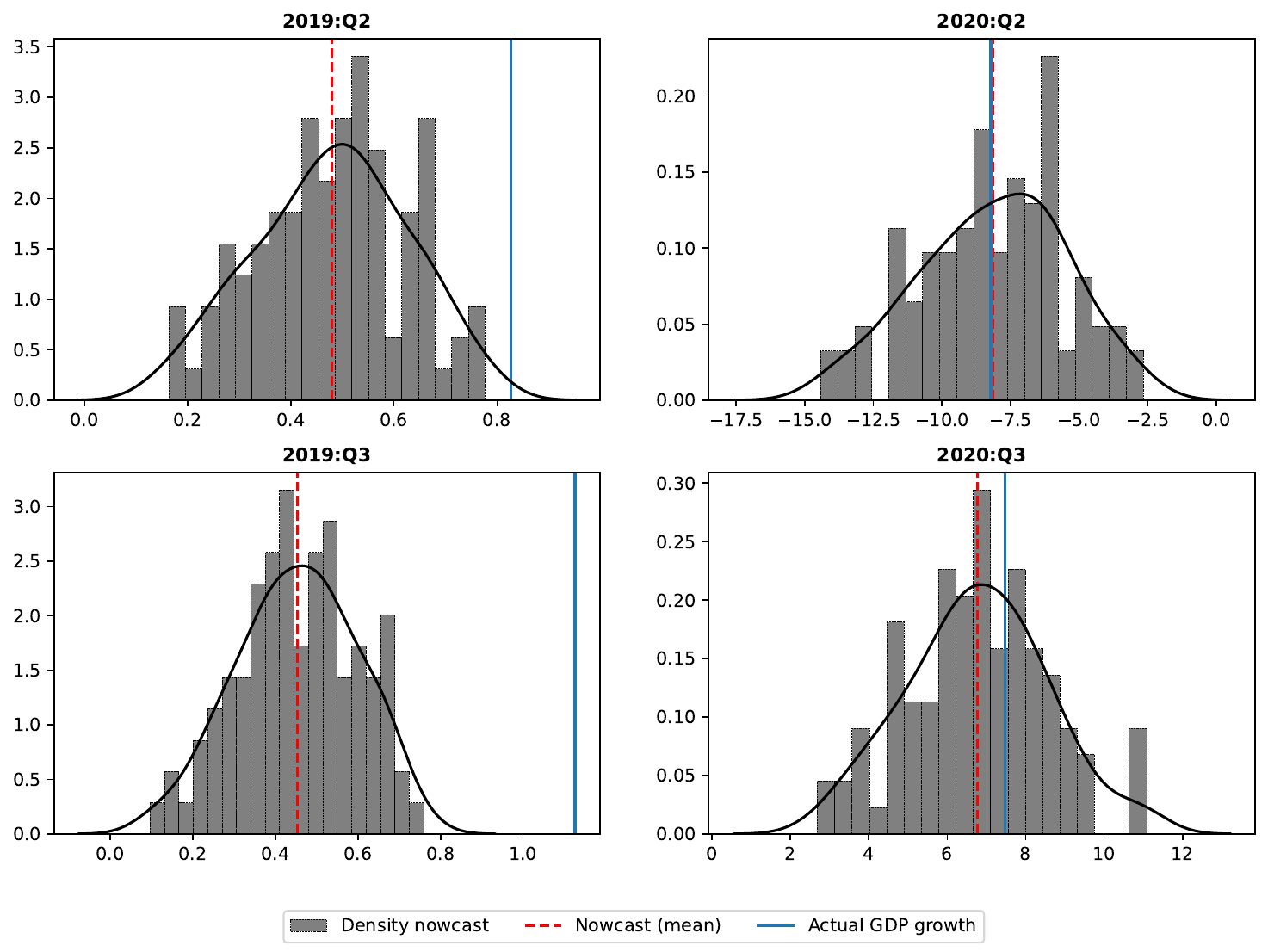}
\caption{Predictive nowcast distributions generated by Bayes by Backprop. 
Nowcasts are conducted based on the 3-month information set ($\Omega_{m_{3}}$), i.e., at the end of the nowcasting window. 
Source: Own calculations based on FRED-MD.} 
\label{fig:bbb_density}
\end{center}
\end{figure}

\subsubsection{Results for Monte Carlo dropout}

Figure \ref{fig:mcdropout_interval} below shows the interval nowcasts generated by Monte Carlo dropout. 
Similarly to \ref{fig:bbb_interval}, Figure \ref{fig:mcdropout_interval} also plots a series of boxplots illustrating how the empirical predictive distribution has evolved during the evaluation period. 

\begin{figure}[H]
\begin{center}
\includegraphics[width = \textwidth]{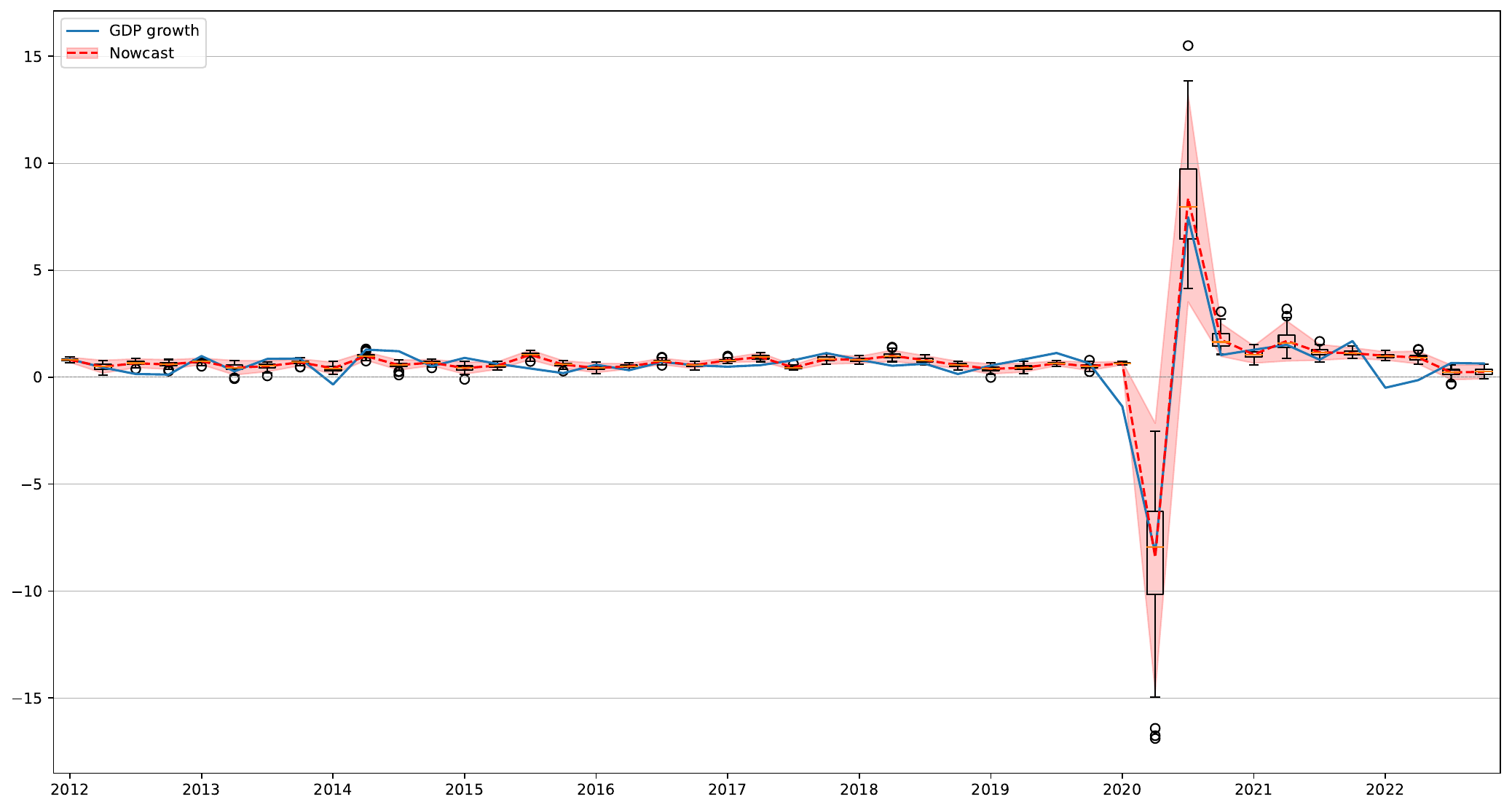}
\caption{Interval nowcasts generated by Monte Carlo dropout. 
To create prediction intervals, we measured two times the standard deviation of the predicitve distribution around the mean (red dashed line). 
Source: Own calculations based on FRED-MD.}
\label{fig:mcdropout_interval}
\end{center}
\end{figure}

Figure \ref{fig:mcdropout_interval} shows that interval nowcasts generated by Monte Carlo dropout are similar to what we have seen previously for Bayes by Backprop in Figure \ref{fig:bbb_interval}. 
The two standard deviation prediction interval is again much narrower compared to DFM during the pre-COVID part of the evaluation period. 
As a result, actual GDP growth falls outside the prediction range in several quarters. 
However, Monte Carlo dropout handles the extreme values of the COVID recession much better and more adaptively than the benchmark DFM. 
As Figure \ref{fig:mcdropout_interval} indicates, the uncertainty range widens considerably during and after the COVID crisis. 
Similar to Bayes by Backprop, the prediction interval shrinks back to its original size by 2022, signaling that economic activity has reverted to its normal state. 
Figure \ref{fig:mcdropout_density} below shows the empirical nowcast distribution (i.e., histograms) generated by Monte Carlo dropout for four different quarters: 
2019:Q2 (upper left), 2019:Q3 (lower-left), 2020:Q2 (upper right) and 2020:Q3 (lower-right). 

\begin{figure}[H]
\begin{center}
\includegraphics[width = \textwidth]{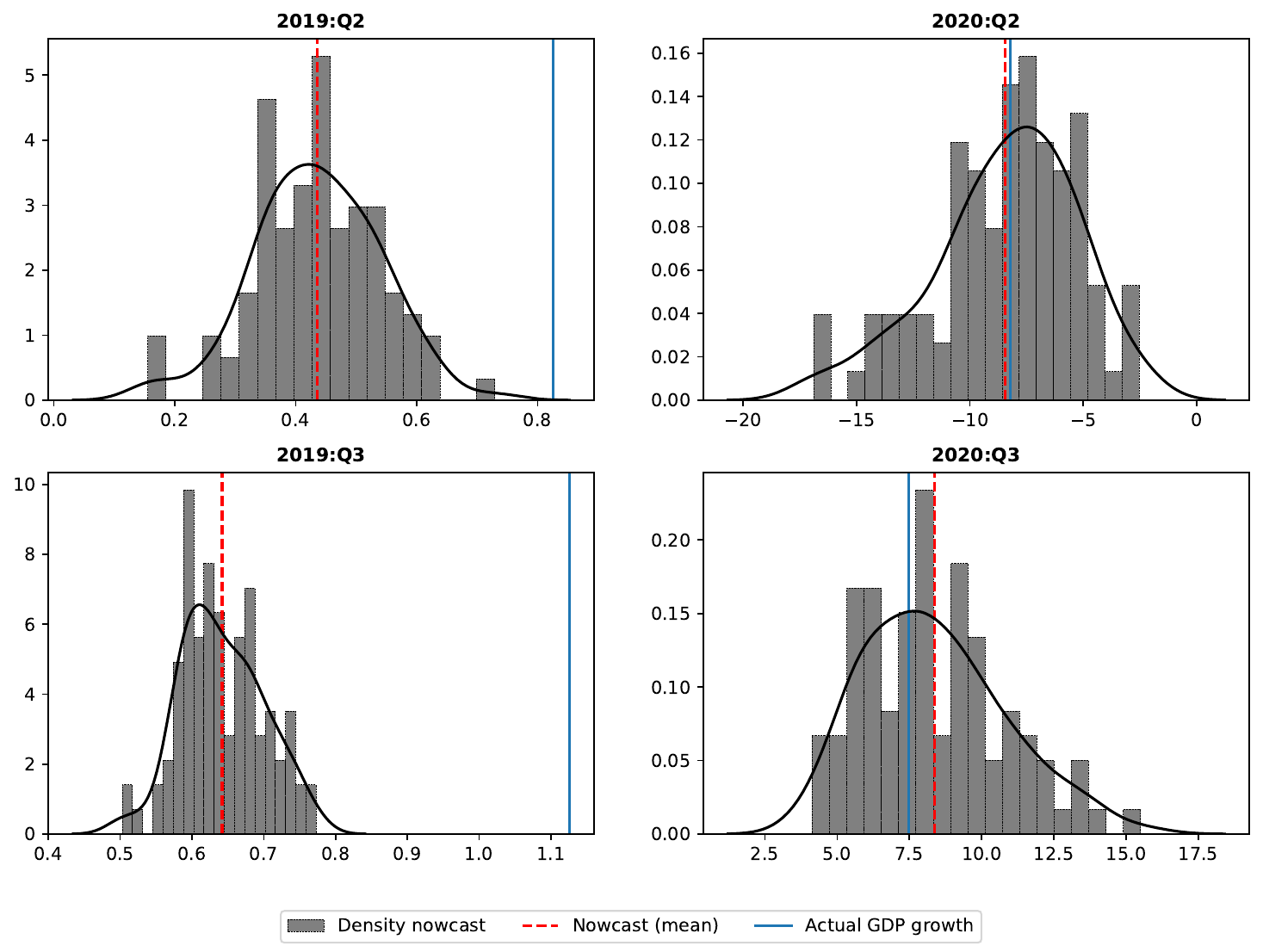}
\caption{Predictive nowcast distributions generated by Monte Carlo dropout.  
Nowcasts are conducted based on the 3-month information set ($\Omega_{m_{3}}$), i.e., at the end of the nowcasting window. 
Source: Own calculations based on FRED-MD.} 
\label{fig:mcdropout_density}
\end{center}
\end{figure}

Similar to Bayes by Backprop, Figure \ref{fig:mcdropout_density} shows that Monte Carlo dropout also learns the behavior of a stochastic volatility model. 
Furthermore, density nowcasts generated by Monte Carlo dropout associate downturns (2020:Q2) with significant negative skew and recovery (2020:Q3) with significant positive skew.

\subsection{Evaluation along the nowcasting window}

\begin{figure}[H]
\begin{center}
\includegraphics[width = \textwidth]{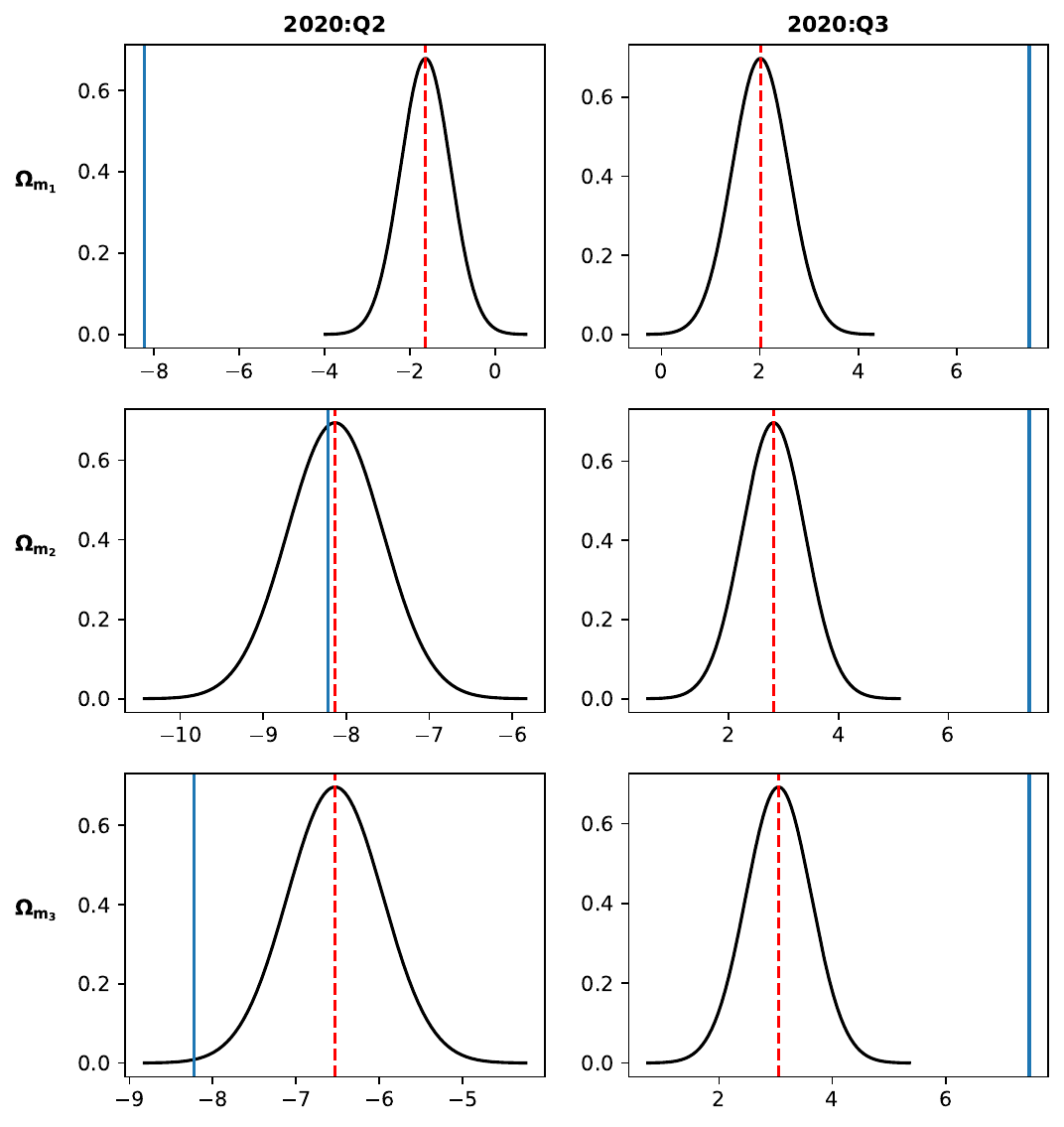}
\caption{Density nowcasts generated by the DFM. 
Nowcasts are conducted based on different intra-quarterly information sets, i.e., at consecutive steps of the nowcasting window. 
Source: Own editing based on FRED-MD.}
\label{fig:dfm_nowcasting-window}
\end{center}
\end{figure}

\begin{figure}[H]
\begin{center}
\includegraphics[width = \textwidth]{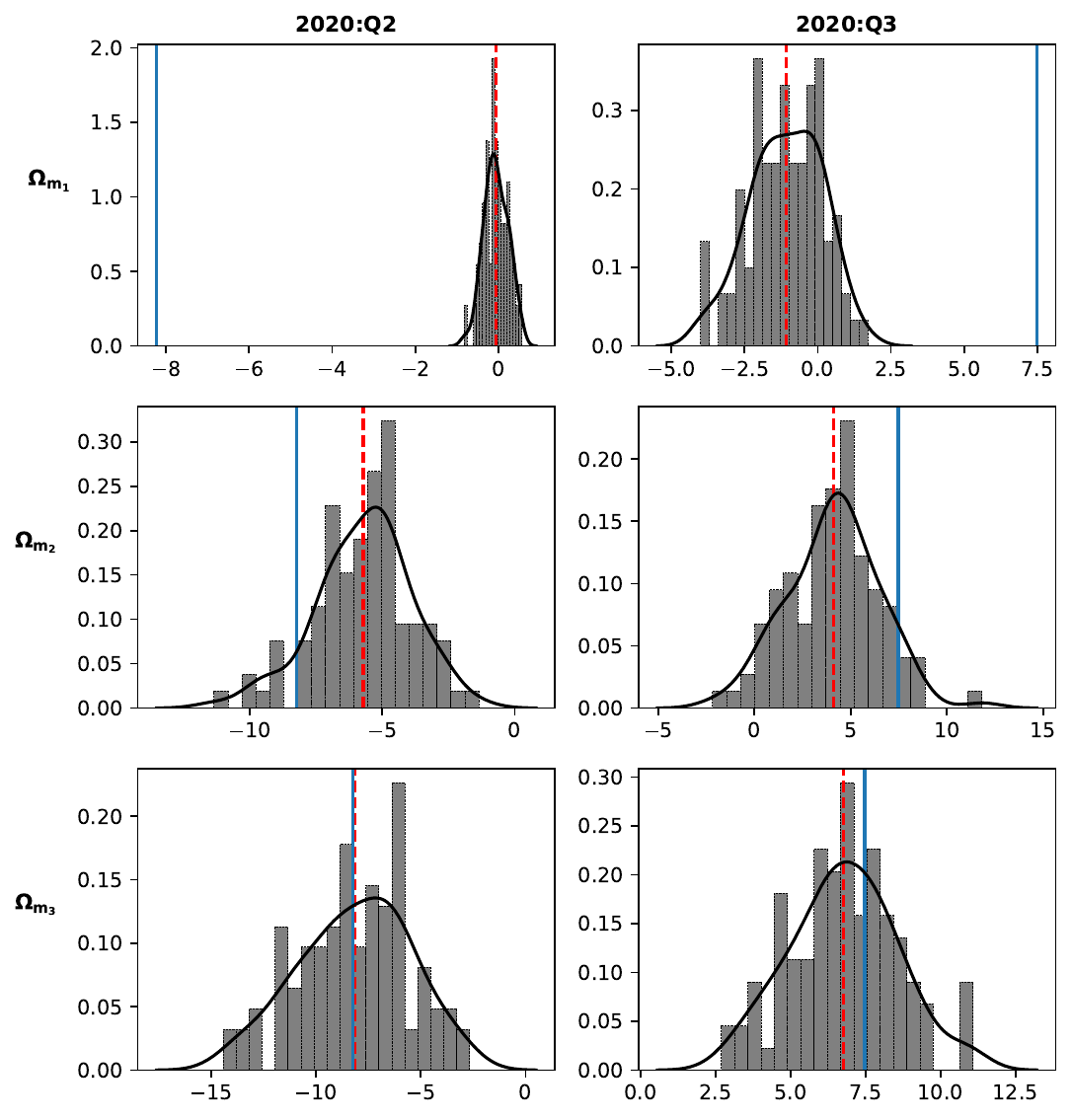}
\caption{Density nowcasts generated by Bayes by Backprop. 
Nowcasts are conducted based on different intra-quarterly information sets, i.e., at consecutive steps of the nowcasting window. 
Source: Own calculations based on FRED-MD.}
\label{fig:bbb_nowcasting-window}
\end{center}
\end{figure}

\newpage

\begin{table}[H]
\centering
\caption{Descriptive statistics of density nowcasts generated by 1D CNN using Bayes by Backprop. 
Nowcasts are generated based on different intra-quarterly information sets, i.e., at consecutive steps of the nowcasting window.}
\begin{tabular}{l L{4cm}| C{3cm}| C{3cm}}
\toprule
             &             &  \textbf{2020:Q2} &  \textbf{2020:Q3} \\
Information set & Statistics &             &             \\
\midrule
             & \textbf{Actual GDP growth} &   $\mathbf{-8.220}$ &    $\mathbf{7.473}$ \\
             & Mean &   $-0.061$ &   $-1.073$ \\
             & Median &   $-0.062$ &   $-1.060$ \\
$\mathbf{\Omega}_{m_{1}}$ & Standard dev. &    $0.287$ &    $1.232$ \\
             & Skew &   $-0.044$ &   $-0.240$ \\
             & Kurtosis &   $-0.296$ &   $-0.388$ \\
             & Jarque-Bera test &    $0.516$ &    $1.698$ \\
\hline
             & \textbf{Actual GDP growth} &   $\mathbf{-8.220}$ &    $\mathbf{7.473}$ \\
             & Mean &   $-5.713$ &    $4.101$ \\
             & Median &   $-5.495$ &    $4.257$ \\
$\mathbf{\Omega}_{m_{2}}$ & Standard dev. &    $1.809$ &    $2.389$ \\
             & Skew &   $-0.423^{*}$ &    $0.018$ \\
             & Kurtosis &    $0.522$ &    $0.422$ \\
             & Jarque-Bera test &    $3.684$ &    $0.493$ \\
\hline
              & \textbf{Actual GDP growth} &   $\mathbf{-8.220}$ &    $\mathbf{7.473}$ \\
             & Mean &   $-8.144$ &    $6.778$ \\
             & Median &   $-7.884$ &    $6.785$ \\
$\mathbf{\Omega}_{m_{3}}$ & Standard dev. &    $2.656$ &    $1.783$ \\
             & Skew &   $-0.219$ &    $0.061$ \\
             & Kurtosis &   $-0.480$ &   $-0.160$ \\
             & Jarque-Bera test &    $1.881$ &    $0.247$ \\
\bottomrule
\end{tabular}
\caption*{\begin{footnotesize}
\textbf{Notes:} This table reports the most important descriptive statistics of density nowcasts generated by Bayes by Backprop. 
The underlying ANN architecture is a 1D CNN. 
Nowcasts are generated at consecutive steps of the nowcasting window. 
Descriptive statistics of the nowcasts are calculated on a sample of 100 elements. 
We use a normalizing factor $N-1$ (i.e., unbiased estimators) for the standard deviation, the skew, and the kurtosis. 
Kurtosis is obtained based on Fisher’s definition of kurtosis (kurtosis of normal = $0$).
The normality of the predictive distribution is checked by the Jarque-Bera test \citep{jarque1980efficient}. 
The stars denote statistical significance at the 10\%($^{*}$), 5\%($^{**}$) and 1\%($^{***}$) level respectively. 
\end{footnotesize}}
\end{table}

\newpage

\begin{figure}[H]
\begin{center}
\includegraphics[width = \textwidth]{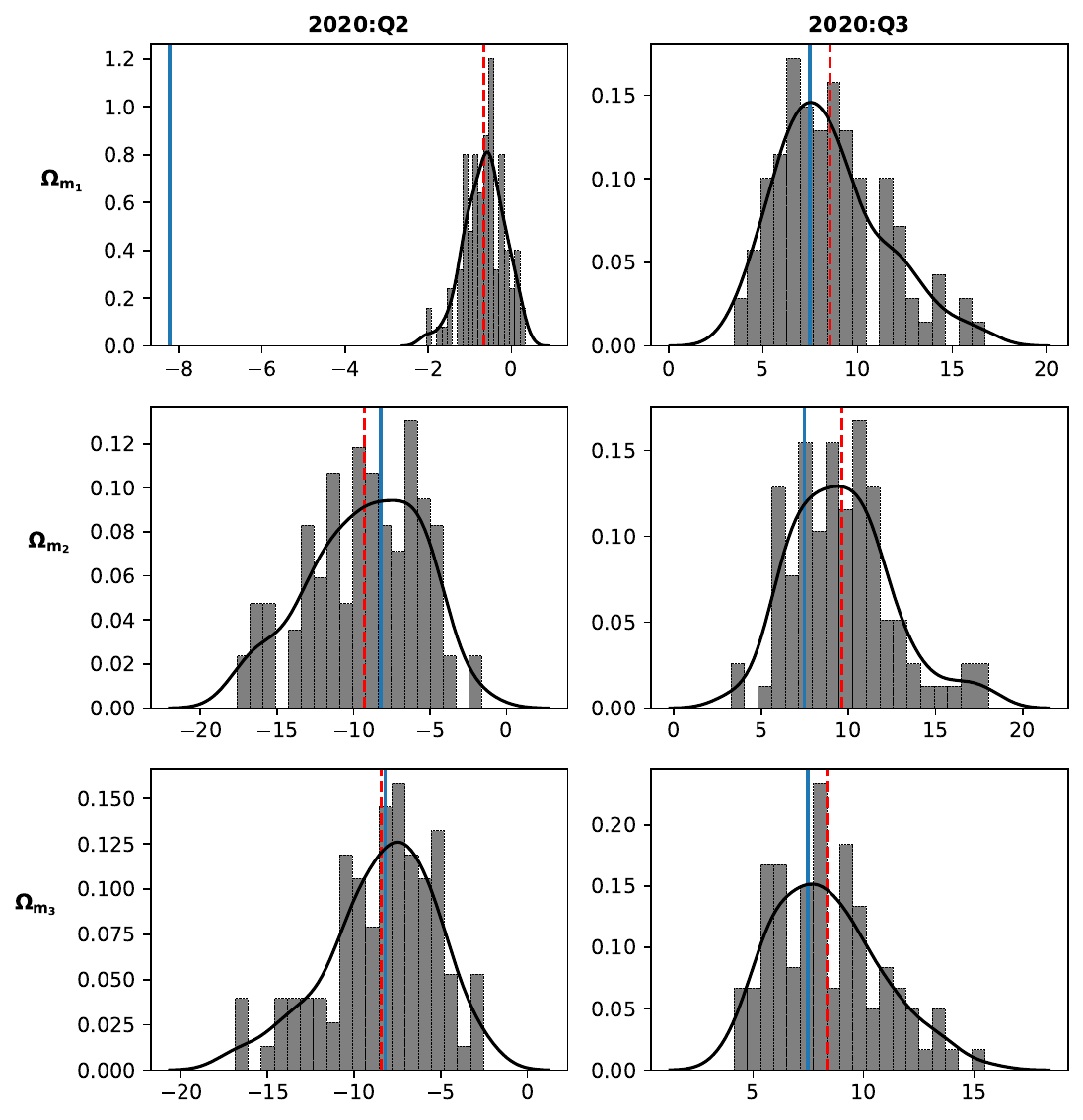}
\caption{Density nowcasts generated by Monte Carlo dropout. 
Nowcasts are conducted based on different intra-quarterly information sets, i.e., at consecutive steps of the nowcasting window. 
Source: Own editing based on FRED-MD.}
\label{fig:mcdropout_nowcasting-window}
\end{center}
\end{figure}

\newpage

\begin{table}[H]
\centering
\caption{Descriptive statistics of density nowcasts generated by Monte Carlo dropout. 
Nowcasts are conducted based on different intra-quarterly information sets, i.e., at consecutive steps of the nowcasting window.}
\begin{tabular}{l L{4cm}| C{3cm}| C{3cm}}
\toprule
             &             &  \textbf{2020:Q2} &  \textbf{2020:Q3} \\
Information set & Statistics &             &             \\
\midrule
             & \textbf{Actual GDP growth} &   $\mathbf{-8.220}$ &    $\mathbf{7.473}$ \\
             & Mean &   $-0.657$ &    $8.557$ \\
             & Median &   $-0.580$ &    $8.150$ \\
$\mathbf{\Omega}_{m_{1}}$ & Standard dev. &    $0.480$ &    $2.831$ \\
             & Skew &   $-0.428^{*}$ &    $0.718^{***}$ \\
             & Kurtosis &    $0.282$ &    $0.171$ \\
             & Jarque-Bera test &    $3.140$ &    $8.390^{**}$ \\
\hline
             & \textbf{Actual GDP growth} &   $\mathbf{-8.220}$ &    $\mathbf{7.473}$ \\
             & Mean &   $-9.284$ &    $9.621$ \\
             & Median &   $-8.813$ &    $9.350$ \\
$\mathbf{\Omega}_{m_{2}}$ & Standard dev. &    $3.669$ &    $2.923$ \\
             & Skew &   $-0.356$ &    $0.645^{***}$ \\
             & Kurtosis &   $-0.534$ &    $0.606$ \\
             & Jarque-Bera test &    $3.384$ &    $7.833^{**}$ \\
\hline
             & \textbf{Actual GDP growth} &   $\mathbf{-8.220}$ &    $\mathbf{7.473}$ \\
             & Mean &   $-8.433$ &    $8.359$ \\
             & Median &   $-7.938$ &    $7.959$ \\
$\mathbf{\Omega}_{m_{3}}$ & Standard dev. &    $3.153$ &    $2.433$ \\
             & Skew &   $-0.597^{**}$ &    $0.566^{**}$ \\
             & Kurtosis &    $0.160$ &   $-0.093$ \\
             & Jarque-Bera test &    $5.796^{*}$ &    $5.274^{*}$ \\
\bottomrule
\end{tabular}
\caption*{\begin{footnotesize}
\textbf{Notes:} This table reports the most important descriptive statistics of density nowcasts generated by Monte Carlo dropout. 
The underlying ANN architecture is a 1D CNN. 
Nowcasts are generated at consecutive steps of the nowcasting window. 
Descriptive statistics of the nowcasts are calculated on a sample of 100 elements. 
We use a normalizing factor $N-1$ (i.e., unbiased estimators) for the standard deviation, the skew, and the kurtosis. 
Kurtosis is obtained based on Fisher’s definition of kurtosis (kurtosis of normal = $0$).
The normality of the predictive distribution is checked by the Jarque-Bera test \citep{jarque1980efficient}. 
The stars denote statistical significance at the 10\%($^{*}$), 5\%($^{**}$) and 1\%($^{***}$) level respectively. 
\end{footnotesize}}
\end{table}

\section{Conclusion}    \label{sec:conclusion}

Our study presents two novel deep learning algorithms, Bayes by Backprop and Monte Carlo dropout, that can generate density nowcasts for US GDP growth. 
As we have seen, these algorithms augment the underlying ANN's default training procedure (i.e., backpropagation), capturing the uncertainty associated with its predictions. 
In our empirical analysis, we use a 1D CNN as the underlying ANN architecture for both algorithms. 
The accuracy of the nowcasts is evaluated based on the mean and the median of the empirical predictive distributions. 
The evaluation (test) period ranges from 2012:Q1 to 2022:Q4 where the DL algorithms generate highly accurate point nowcasts compared to a naive constant growth model for GDP and a benchmark DFM specification. 

Assessing the (model) uncertainty associated with the nowcasts, we use the standard deviation of the predictive distributions. 
Prediction intervals for the DFM work as expected: 
Actual GDP growth falls within the prediction interval according to the probability determined by the underlying model. 
Unfortunately though, prediction intervals produced by the benchmark DFM miss the actuals when it matters most: in times of severe economic downturn. 
By contrast, the DL algorithms adjust the scale (dispersion) of the predictive distributions dynamically, learning the behavior of a stochastic volatility model. 
By doing so, they capture the sudden change in the economic environment during the onset of the COVID crisis. 
Density nowcasts produced by these algorithms also indicate that uncertainty remained persistently higher in the post-COVID part of the evaluation period. 

In general, predictive densities obtained from these models are not significantly different from a normal distribution. 
Along with that, both algorithms are able to dynamically adjust the location (mean), scale (variance) and shape (skew) of the empirical predictive distribution. 
From the perspective of the policy-makers, those features can deliver additional insights into the current economic state. 
Density nowcasts generated by Monte Carlo dropout associate downturns (2020:Q2) with significant negative skew and recovery (2020:Q3) with significant positive skew. 
Similarly, Bayes by backprop produces a density nowcast for 2020:Q2 which shows significant negative skew at the second step of the nowcasting window (i.e., in May). 
Asymmetry of the predictive distributions, indicated by a significant skew towards negative or positive values, can provide valuable additional information, signaling the expected direction of the prediction error. 
The results indicate that both Bayes by Backprop and Monte Carlo dropout can effectively augment the scope and functionality of ANNs, rendering them a fully compatible and competitive alternative for classical time series approaches.

\section*{Acknowledgements}  \label{sec:acknowledgements}

We would like to thank Róbert Lieli, professor at the Central European University, for his invaluable feedback and support. 
He has helped and inspired our work in many ways.

\medskip
\printbibliography

\newpage

\appendix
\section{Data appendix}     \label{app:data}

The TCODE column denotes the following data transformation for a series $x$:
\begin{enumerate}
\item No transformation
\item $\Delta x_t$
\item $\Delta^{2} x_t$
\item $log(x_t)$
\item $\Delta log(x_t)$
\item $\Delta^{2} log(x_t)$
\item $\Delta \left( \frac{x_t - x_{t-1}}{x_{t-1}} \right)$
\end{enumerate} 

The FRED column gives mnemonics in FRED followed by a short description. 
Some series require adjustments to the raw data available in FRED. 
These variables are tagged by an asterisk to indicate that they have been adjusted and thus differ from the series from the source. 
For a detailed summary of the adjustments see \cite{mccracken2016fred}. 

\vspace{30pt}

\begin{table}[H]
\centering
\caption*{Group 1: Output and income}
\begin{tabular}{lllll}
\toprule
  & ID & tcode & FRED & Description \\
\midrule
1 & 1 & 5 & RPI & Real Personal Income \\
2 & 2 & 5 & W875RX1 & Real personal income ex transfer receipts \\
3 & 6 & 5 & INDPRO & IP Index \\
4 & 7 & 5 & IPFPNSS & IP: Financial Products and Nonindustrial Supplies \\
5 & 8 & 5 & IPFINAL & IP: Final Products (Market Group) \\
6 & 9 & 5 & IPCONGD & IP: Consumer Goods \\
7 & 10 & 5 & IPDCONGD & IP: Durable Consumer Goods \\
8 & 11 & 5 & IPNCONGD & IP: Nondurable Consumer Goods \\
9 & 12 & 5 & IPBUSEQ & IP: Business Equipment \\
10 & 13 & 5 & IPMAT & IP: Materials \\
11 & 14 & 5 & IPDMAT & IP: Durable Materials \\
12 & 15 & 5 & IPNMAT & IP: Nondurable Materials \\
13 & 16 & 5 & IPMANSICS & IP: Manufacturing (SIC) \\
14 & 17 & 5 & IPB51222s & IP: Residential Utilities \\
15 & 18 & 5 & IPFUELS & IP: Fuels \\
16 & 19 & 1 & NAPMPI & ISM Manufacturing: Production Index \\
17 & 20 & 2 & CUMFNS & Capacity Utilization: Manufacturing \\
\bottomrule
\end{tabular}
\end{table}

\newpage

\begin{table}[H]
\centering
\caption*{Group 2: Labor market}
\begin{tabular}{lllll}
\toprule
  & ID & tcode & FRED & Description \\
\midrule
1 & 21* & 2 & HWI & Help-Wanted Index for United States \\
2 & 22* & 2 & HWIURATIO & Ratio of Help Wanted/No. Unemployed \\
3 & 23 & 5 & CLF160OV & Civilian Labor Force \\
4 & 24 & 5 & CE160V & Civilian Employment \\
5 & 25 & 2 & UNRATE & Civilian Unemployment Rate \\
6 & 26 & 2 & UEMPMEAN & Average Duration of Unemployment (Weeks) \\
7 & 27 & 5 & UEMPLT5 & Civilians Unemployed -- Less Than 5 Weeks \\
8 & 28 & 5 & UEMP5TO14 & Civilians Unemployed for 5-14 Weeks \\
9 & 29 & 5 & UEMP15OV & Civilians Unemployed -- 15 Weeks and Over \\
10 & 30 & 5 & UEMP15T26 & Civilians Unemployed for 15 -- 26 Weeks \\
11 & 31 & 5 & UEMP27OV & Civilians Unemployed for 27 Weeks and Over \\
12 & 32* & 5 & CLAIMSx & Initial Claims \\
13 & 33 & 5 & PAYEMS & All Employees: Total nonfarm \\
14 & 34 & 5 & USGOOD & All Employees: Goods-Producing Industries \\
15 & 35 & 5 & CES1021000001 & All Employees: Mining and Logging: Industries \\
16 & 36 & 5 & USCONS & All Employees: Construction \\
17 & 37 & 5 & MANEMP & All Employees: Manufacturing \\
18 & 38 & 5 & DMANEMP & All Employees: Durable Goods \\
19 & 39 & 5 & NDMANEMP & All Employees: Nondurable Goods \\
20 & 40 & 5 & SRVPRD & All Employees: Service-Providing Industries \\
21 & 41 & 5 & USTPU & All Employees: Trade, Transportation and Utilities \\
22 & 42 & 5 & USWTRADE & All Employees: Wholesale Trade \\
23 & 43 & 5 & USTRADE & All Employees: Retail Trade \\
24 & 44 & 5 & USFIRE & All Employees: Financial Activities \\
25 & 45 & 5 & USGOVT & All Employees: Government \\
26 & 46 & 1 & CES0600000007 & Avg Weekly Hours: Goods-Producing \\
27 & 47 & 2 & AWOTMAN & Avg Weekly Overtime Hours: Manufacturing \\
28 & 48 & 1 & AWHMAN & Avg Weekly Hours: Manufacturing \\
29 & 49 & 1 & NAPMEI & ISM Manufacturing: Employment Index \\
30 & 127 & 6 & CES0600000008 & Avg Hourly Earnings: Goods-Producing \\
31 & 128 & 6 & CES2000000008 & Avg Hourly Earnings: Construction \\
32 & 129 & 6 & CES3000000008 & Avg Hourly Earnings: Manufacturing \\
\bottomrule
\end{tabular}
\end{table}

\newpage

\begin{table}[H]
\centering
\caption*{Group 3: Housing}
\begin{tabular}{lllll}
\toprule
  & ID & tcode & FRED & Description \\
\midrule
1 & 50 & 4 & HOUST & Housing Starts: Total New Privately Owned \\
2 & 51 & 4 & HOUSTNE & Housing Starts: Northeast \\
3 & 52 & 4 & HOUSTMW & Housing Starts: Midwest \\
4 & 53 & 4 & HOUSTS & Housing Starts: South \\
5 & 54 & 4 & HOUSTW & Housing Starts: West \\
6 & 55 & 4 & PERMIT & New Private Housing Permits (SAAR) \\
7 & 56 & 4 & PERMITNE & New Private Housing Permits: Northeast (SAAR) \\
8 & 57 & 4 & PERMITMW & New Private Housing Permits: Midwest (SAAR) \\
9 & 58 & 4 & PERMITS & New Private Housing Permits: South (SAAR) \\
10 & 59 & 4 & PERMITW & New Private Housing Permits: West (SAAR) \\
\bottomrule
\end{tabular}
\end{table}

\vspace{30pt}

\begin{table}[H]
\centering
\caption*{Group 4: Consumption, orders and inventories}
\begin{tabular}{lllll}
\toprule
  & ID & tcode & FRED & Description \\
\midrule
1 & 3 & 5 & DPCERA3M086SBEA & Real personal consumption expenditures \\
2 & 4* & 5 & CMRMTSPLx & Real Manu. and Trade Industries Sales \\
3 & 5* & 5 & RETAILx & Retail and Food Services Sales \\
4 & 60 & 1 & NAPM & ISM: PMI Composite Index \\
5 & 61 & 1 & NAPMNOI & ISM: New Orders Index \\
6 & 62 & 1 & NAPMSDI & ISM: Supplier Deliveries Index \\
7 & 63 & 1 & NAPMII & ISM: Inventories Index \\
8 & 64 & 5 & ACOGNO & New Orders for Consumer Goods \\
9 & 65* & 5 & AMDMNOx & New Orders for Durable Goods \\
10 & 66* & 5 & ANDENOx & New Orders for Nondefense Capital Goods \\
11 & 67* & 5 & AMDMUOx & Unfilled Orders for Durable Goods \\
12 & 68* & 5 & BUSINVx & Total Business Inventories \\
13 & 69* & 2 & ISRATIOx & Total Business: Inventories to Sales Ratio \\
14 & 130* & 2 & UMSCENTx & Consumer Sentiment Index \\
\bottomrule
\end{tabular}
\end{table}

\newpage

\begin{table}[H]
\centering
\caption*{Group 5: Money and credit}
\begin{tabular}{lllll}
\toprule
  & ID & tcode & FRED & Description \\
\midrule
1 & 70 & 6 & M1SL & M1 Money Stock \\
2 & 71 & 6 & M2SL & M2 Money Stock \\
3 & 72 & 5 & M2REAL & Real M2 Money Stock \\
4 & 73 & 6 & AMBSL & St. Louis Adjusted Monetary Base \\
5 & 74 & 6 & TOTRESNS & Total Reserves of Depository Institutions \\
6 & 75 & 7 & NONBORRES & Reserves of Depository Institutions \\
7 & 76 & 6 & BUSLOANS & Commercial and Industrial Loans \\
8 & 77 & 6 & REALLN & Real Estate Loans at All Commercial Banks \\
9 & 78 & 6 & NONREVSL & Total Nonrevolving Credit \\
10 & 79* & 2 & CONSPI & Nonrevolving consumer credit to Personal Income \\
11 & 131 & 6 & MZMSL & MZM Money Stock \\
12 & 132 & 6 & DTCOLNVHFNM & Consumer Motor Vehicle Loans Outstanding \\
13 & 133 & 6 & DTCTHFNM & Total Consumer Loans and Leases Outstanding \\
14 & 134 & 6 & INVEST & Securities in Bank Credit at All Commercial Banks \\
\bottomrule
\end{tabular}
\end{table}

\newpage

\begin{table}[H]
\centering
\caption*{Group 6: Interest and exchange rates}
\begin{tabular}{lllll}
\toprule
  & ID & tcode & FRED & Description \\
\midrule
1 & 84 & 2 & FEDFUNDS & Effective Federal Funds Rate \\
2 & 85* & 2 & CP3Mx & 3-Month AA Financial Commercial Paper Rate \\
3 & 86 & 2 & TB3MS & 3-Month Treasury Bill \\
4 & 87 & 2 & TB6MS & 6-Month Treasury Bill \\
5 & 88 & 2 & GS1 & 1-Year Treasury Rate \\
6 & 89 & 2 & GS5 & 5-Year Treasury Rate \\
7 & 90 & 2 & GS10 & 10-Year Treasury Rate \\
8 & 91 & 2 & AAA & Moody's Seasoned Aaa Corporate Bond Yield \\
9 & 92 & 2 & BAA & Moody's Seasoned Baa Corporate Bond Yield \\
10 & 93* & 1 & COMPAPFFx & 3-Month Commercial Paper Minus FEDFUNDS \\
11 & 94 & 1 & TB3SMFFM & 3-Month Treasury C Minus FEDFUNDS \\
12 & 95 & 1 & TB6SMFFM & 6-Month Treasury C Minus FEDFUNDS \\
13 & 96 & 1 & T1YFFM & 1-Year Treasury C Minus FEDFUNDS \\
14 & 97 & 1 & T5YFFM & 5-Year Treasury C Minus FEDFUNDS \\
15 & 98 & 1 & T10YFFM & 10-Year Treasury C Minus FEDFUNDS \\
16 & 99 & 1 & AAAFFM & Moody's Aaa Corporate Bond Minus FEDFUNDS \\
17 & 100 & 1 & BAAFFM & Moody's Baa Corporate Bond Minus FEDFUNDS \\
18 & 101 & 5 & TWEXMMTH & Trade Weighted U.S. Dollar Index: Major Currencies \\
19 & 102* & 5 & EXSZUSx & Switzerland / U.S. Foreign Exchange Rate \\
20 & 103* & 5 & EXJPUSx & Japan / U.S. Foreign Exchange Rate \\
21 & 104* & 5 & EXUSUKx & U.S. / U.K. Foreign Exchange Rate \\
22 & 105* & 5 & EXCAUSx & Canada / U.S. Foreign Exchange Rate \\
\bottomrule
\end{tabular}
\end{table}

\newpage

\begin{table}[H]
\centering
\caption*{Group 7: Prices}
\begin{tabular}{lllll}
\toprule
  & ID & tcode & FRED & Description \\
\midrule
1 & 106 & 6 & WPSFD49207 & PPI: Finished Goods \\
2 & 107 & 6 & WPSFD49502 & PPI: Finished Consumer Goods \\
3 & 108 & 6 & WPSID61 & PPI: Intermediate Materials \\
4 & 109 & 6 & WPSID62 & PPI: Crude Materials \\
5 & 110* & 6 & OILPRICEx & Crude Oil, spliced WTI and Cushing \\
6 & 111 & 6 & PPICMM & PPI: Metals and metal products \\
7 & 112 & 1 & NAPMPRI & ISM Manufacturing: Prices Index \\
8 & 113 & 6 & CPIAUCSL & CPI: All Items \\
9 & 114 & 6 & CPIAPPSL & CPI: Apparel \\
10 & 115 & 6 & CPITRNSL & CPI: Transportation \\
11 & 116 & 6 & CPIMEDSL & CPI: Medical Care \\
12 & 117 & 6 & CUSR0000SAC & CPI: Commodities \\
13 & 118 & 6 & CUSR0000SAD & CPI: Durables \\
14 & 119 & 6 & CUSR0000SAS & CPI: Service \\
15 & 120 & 6 & CPIULFSL & CPI: All Items less Food \\
16 & 121 & 6 & CUSR0000SA0L2 & CPI: All Items less Shelter \\
17 & 122 & 6 & CUSR0000SA0L5 & CPI: All Items less Medical Care \\
18 & 123 & 6 & PCEPI & Personal Cons. Expend.: Chain Index \\
19 & 124 & 6 & DDURRG3M086SBEA & Personal Cons. Expend.: Durable Goods \\
20 & 125 & 6 & DNDGRG3M086SBEA & Personal Cons. Expend.: Nondurable Goods \\
21 & 126 & 6 & DSERRG3M086SBEA & Personal Cons. Expend.: Services \\
\bottomrule
\end{tabular}
\end{table}

\vspace{20pt}

\begin{table}[H]
\centering
\caption*{Group 8: Stock market}
\begin{tabular}{lllll}
\toprule
  & ID & tcode & FRED & Description \\
\midrule
1 & 80* & 5 & S\&P 500 & S\&P's Common Stock Price Index: Composite \\
2 & 81* & 5 & S\&P: indust & S\&P's Common Stock Price Index: Industrials \\
3 & 82* & 2 & S\&P div yield & S\&P's Composite Common Stock: Dividend Yield \\
4 & 83* & 5 & S\&P PE ratio & S\&P's Composite Common Stock: Price-Earnings Ratio \\
\bottomrule
\end{tabular}
\end{table}

\end{document}